\documentclass[10pt,conference, review]{IEEEtran}
\usepackage{cite}
\usepackage{amsmath,amssymb,amsfonts}
\usepackage{graphicx}
\usepackage{enumitem}
\usepackage{caption,subcaption}
\usepackage{textcomp}
\usepackage{xcolor}
\usepackage[linesnumbered,commentsnumbered,ruled,vlined]{algorithm2e}

\usepackage{stmaryrd}

\usepackage{booktabs}   

\usepackage{amsthm}
\usepackage{tcolorbox}

\theoremstyle{definition}

\newtheorem{example}{Example}
\newtheorem{definition}{{Definition}}
\newtheorem{thm4}{Theorem}
\newtheorem{theorem}[thm4]{Theorem}

\newtheorem{thm8}{Definition}
\newtheorem{lemma}[thm8]{Lemma}

\newcommand{\mybox}[1]{
	\begin{tcolorbox}[
		boxsep=-0.5pt,
		standard jigsaw,
		boxrule=1 pt,
		opacityback=0,
		sharp corners
		]
		#1
	\end{tcolorbox}
}

\usepackage{hyperref}
\PassOptionsToPackage{colorlinks, dvipsnames}{xcolor}
\hypersetup{
	colorlinks=true,
	linkcolor=blue,
	citecolor=violet,
	filecolor=magenta, 
	urlcolor=blue,
}

\usepackage{semantic}
\usepackage[utf8]{inputenc}
\usepackage{cleveref}

\usepackage{listings}
\usepackage{lipsum}                     

\usepackage{xargs}                      

\usepackage{balance}
\usepackage{pifont}
\usepackage{mathtools}
\usepackage{amsmath}
\usepackage{url}

\usepackage{tabulary}
\usepackage{etoolbox}
\usepackage{mathrsfs}  

\usepackage{enumitem}

\usepackage{balance}
\usepackage{multicol}

\graphicspath{ {resources/}{data/} }

\usepackage{mwe} 
\usepackage{pgf,tikz}

\newif\ifshowcomments

\renewcommand{\baselinestretch}{1}

\showcommentstrue

\ifshowcomments
\newcommand{\wcp}[1]{\mytodocyan{[wcp: #1]}}
\newcommand{\yao}[1]{\mytodored{[yao: #1]}}
\newcommand{\gang}[1]{\mytodoblue{[gang: #1]}}
\else
\newcommand{\wcp}[1]{}
\newcommand{\yao}[1]{}
\newcommand{\gang}[1]{}
\fi

\newcommand{\mytodoblue}[1]{\textcolor{blue}{\ding{46}~{\sf}~#1}}
\newcommand{\mytodored}[1]{\textcolor{red}{\ding{46}~{\sf}~#1}}

\newcommand{\mytodocyan}[1]{\textcolor{cyan}{\ding{46}~{\sf}~#1}}

\newcommand{\ToolName}{\textsc{EqDAC}}

\newcommand{\UDP}{\textsc{UDP}}

\newcommand{\EQUITAS}{\textsc{EQUITAS}}
\newcommand{\Green}{\textsc{Green}}
\newcommand{\ToolNameNoInvariant}{\textsc{EqDAC-ND}}
\newcommand{\ToolNameNoCertificate}{\textsc{EqDAC-NI}}
\newcommand{\ToolNameNoSemantics}{\textsc{EqDAC-NS}}

\newcommand{\TotalRuleNumber}{30,801}
\newcommand{\TotalEquivalentPairNumber}{26,789}
\newcommand{\TotalMustEquivalentPairNumber}{25,952}
\newcommand{\TotalLastStageEquivalentPairNumber}{837}
\newcommand{\NoCertificateTime}{4.48}
\newcommand{\TotalEqRuleNumber}{11,538}
\newcommand{\TotalDeletedRuleNumber}{7,842}
\newcommand{\TotalMustDeletedRuleNumber}{7,296}
\newcommand{\TotalLastStageEquivalentNumber}{546}
\newcommand{\FilteringDiffNumber}{37}
\newcommand{\NoSemanticsTime}{2.13}
\newcommand{\TotalTime}{2.89}



\usepackage[switch]{lineno}

\usepackage{quoting}
\usepackage{tabularx}
\usepackage{csquotes}
\usepackage{cleveref}
\crefname{section}{§}{§§}
\Crefname{section}{§}{§§}

\usepackage[symbol]{footmisc}

\definecolor{codegreen}{rgb}{0,0.6,0}
\definecolor{codegray}{rgb}{0.5,0.5,0.5}
\definecolor{codepurple}{rgb}{0.58,0,0.82}
\definecolor{backcolour}{rgb}{0.95,0.95,0.92}

\lstdefinestyle{mystyle}{
	backgroundcolor=\color{white},   
	commentstyle=\color{codepurple},
	language={C}, 
	basicstyle=\ttfamily\scriptsize,
	keywordstyle=\scriptsize\bfseries\color{black},
	keywordstyle = [2]{\scriptsize\bfseries\color{blue}},
	keywordstyle = [3]{\scriptsize\bfseries\color{codepurple}},
	keywordstyle = [4]{\scriptsize\bfseries\color{teal}},
	otherkeywords = {if, else, assert, and, nil, not},
	morekeywords = [2]{abs, capital, contains},
	morekeywords = [3]{0},
	numberstyle=\color{codepurple},
	stringstyle=\color{codepurple},
	breakatwhitespace=false,           
	captionpos=b,
	numbers=none,
	showspaces=false,                
	showstringspaces=false,
	showtabs=false, 
	tabsize=2
}

\begin{document}

	
    \title{Verifying Data Constraint Equivalence \\ in FinTech Systems} 

\author{\IEEEauthorblockN{Chengpeng Wang\textsuperscript{$\ast$}, 
		Gang Fan\textsuperscript{$\dagger$},
		Peisen Yao\textsuperscript{$\ddagger$},
		Fuxiong Pan\textsuperscript{$\dagger$}, and
		Charles Zhang\textsuperscript{$\ast$}}
	\IEEEauthorblockA{\textsuperscript{$\ast$}Department of Computer Science and Engineering\\ 
		The Hong Kong University of Science and Technology, Hong Kong, China.\\
		Email: \{cwangch, charlesz\}@cse.ust.hk,\\
		\textsuperscript{$\dagger$}Ant Group, Shenzhen, China. Email: \{fangang, fuxiong.pfx\}@antgroup.com,\\
		\textsuperscript{$\ddagger$}Zhejiang University, Hangzhou, China. Email: pyaoaa@zju.edu.cn}
} 

	\maketitle

	\begin{abstract}
Data constraints are widely used in FinTech systems
for monitoring data consistency and diagnosing anomalous data manipulations.
However, many equivalent data constraints are created redundantly during the development cycle,
slowing down the FinTech systems and causing unnecessary alerts.
We present \ToolName, an efficient decision procedure to determine the data constraint equivalence.
We first propose the symbolic representation for semantic encoding
and then introduce two light-weighted analyses to refute and prove the equivalence, respectively,
which are proved to achieve in polynomial time.
We evaluate  \ToolName\ upon \TotalRuleNumber\ data constraints in a FinTech system.
It is shown that \ToolName\ detects \TotalEqRuleNumber\ equivalent data constraints in three hours.
It also supports efficient equivalence searching with an average time cost of 1.22 seconds,
enabling the system to check new data constraints upon submission.
\end{abstract}

	\smallskip
	
	\begin{IEEEkeywords}
		Equivalence Verification, Data Constraints, FinTech Systems
	\end{IEEEkeywords}

\maketitle

\section{Introduction}
With the rapid development of E-commerce,
FinTech systems have become increasingly essential to industrial production.
They consist of a cluster of database-backed applications manipulating large amounts of sensitive data~\cite{Li19}.
Any incorrect data value can yield system misbehaviors and cause immeasurable financial losses.
To ensure reliability,
it is a common practice to specify target properties as data constraints~\cite{florez2021empirical, florez2022retrieving}
for runtime checking.
If a data constraint is violated,
developers can receive an alert for further diagnosis.

Unfortunately,
the continuous submissions from developers make data constraints accumulate rapidly
and can even introduce redundancy.
In a global FinTech company A, 103 developers submitted 2,306 data constraints in the first quarter of 2022.
Unaware of previous submissions, they create equivalent data constraints,
which gradually become the technical debt~\cite{ernst2021technical},
wasting computing resources and increasing the burden of system maintenance.
To resolve the redundancy,
the developers expect to search the existing equivalent data constraints before submitting new ones, thereby avoiding redundant submissions.
Besides, quality assurance teams are eager to examine data constraint repositories regularly,
seeking more opportunities for optimization based on the equivalence relation.
Thus, it is relevant to verify the data constraint equivalence for better maintenance 
in a FinTech system.

\smallskip
\textbf{Goal and Challenges.}
We aim to design a decision procedure determining whether two data constraints are equivalent.
However, it is stunningly challenging to find a solution fitting industrial requirements.
First, the decision procedure should be highly efficient,
as FinTech systems often contain tens of thousands of data constraints,
which amplify the efficiency bottleneck greatly.
Any inefficiency in the decision procedure can result in significant burdens of adoption.
Second, it is crucial to guarantee soundness and prove the equivalence as completely as possible.
Otherwise, it would remove necessary data constraints or miss equivalent ones, resulting in financial losses or hiding opportunities for further optimization, respectively.
In reality, data constraints contain various operations upon different data types,
increasing the difficulty of achieving these objectives simultaneously.

\smallskip
\textbf{Existing Efforts.}
There have been two lines of research on equivalence verification.
One line of the techniques leverages the specified rewrite rules 
and checks whether a program can be transformed to the other via \emph{term rewriting}~\cite{Letelier0PS13, Chu17HoTTSQL}.
Although the rewrite rules theoretically ensure soundness,
they can only identify restrictive forms of equivalent patterns~\cite{Necula00},
and the vast search space of applying rewrite rules also brings great overhead~\cite{Max21Egg}.
The other line encodes the program semantics with logical formulas and performs the symbolic reasoning by invoking an SMT solver~\cite{Zhou19VLDB, zhou2020symbolic, BadihiA0R20}.
It provides a general approach to verify the equivalence,
while an SMT solver is not efficient enough to reason a large number of data constraints.
The solver has to be invoked thousands of times in the equivalence clustering and searching,
accumulating the overhead and finally degrading the overall efficiency~\cite{BouchetCCDGHJMP20}.

\smallskip
\textbf{Insight and Solution.}
Our key idea originates from two critical observations.
First, non-equivalent data constraints often contain different variables, literals, or operators.
For example, the data constraint in Fig.~\ref{subfig:motivatingEx1a} examines the attributes \emph{oid} and \emph{in} in the table \emph{t},
while the data constraint in Fig.~\ref{subfig:motivatingEx1b} examines the attributes \emph{iid} and \emph{new} instead.
The lexical differences guide the generation of concrete values to make two data constraints evaluate differently.
Second, equivalent data constraints often converge towards similar syntactic structures.
For instance, the data constraints in Fig.~\ref{subfig:motivatingEx1d} and Fig.~\ref{subfig:motivatingEx1c}
only differ in the orders of assertions, branches, and commutative operands after eliminating user-defined variables.
The isomorphic syntactic structures are the witness of their equivalence.
Thus, we can leverage the lexical differences and syntactic isomorphism to efficiently refute and prove the equivalence, respectively,
avoiding unnecessary SMT solving for better performance.

\begin{figure}[t]
\centering
\begin{subfigure}{0.42\linewidth}
\centering
\begin{lstlisting}[style=mystyle]
s = 'IN';
if(contains(t.ty,s))
	assert(t.in > 0);
else
	assert(t.out > 0);
assert(t.amt > 0);
assert(t.oid != 0);
\end{lstlisting}
\caption{}
\label{subfig:motivatingEx1a}
\end{subfigure}
\begin{subfigure}{0.56\linewidth}
\begin{lstlisting}[style=mystyle]
if(contains(t.ty,'IN')){
	assert(t.old == t.new - t.in);
} else {
	assert(t.old == t.new + t.out);
}
assert(t.oid != 0);
assert(t.iid != 0);
\end{lstlisting}
\caption{}
\label{subfig:motivatingEx1d}
\end{subfigure}

\begin{subfigure}{0.45\linewidth}
\begin{lstlisting}[style=mystyle]
s = 'IN';
if(not contains(t.ty,s))
	assert(t.out > 0);
else
	assert(t.new > 0);
assert(t.amt > 0);
assert(t.iid != 0);
\end{lstlisting}
\caption{}
\label{subfig:motivatingEx1b}
\end{subfigure}
\begin{subfigure}{0.53\linewidth}
\begin{lstlisting}[style=mystyle]
assert(t.iid != 0);
assert(t.oid != 0);
if(not contains(t.ty,'IN'))
  cash = t.out + t.new;
else
  cash = t.new - t.in;
assert(cash == t.old);
\end{lstlisting}
\caption{}
\label{subfig:motivatingEx1c}
\end{subfigure}
\caption{Examples of data constraints}   
\vspace{-3mm}
\label{fig:motivatingEx}
\end{figure}

Based on the insight, we present \ToolName, an efficient decision procedure for the equivalence verification.
We establish a first-order logic (FOL) formula as the symbolic representation to depict the semantics.
To refute the equivalence, we perform the \emph{divergence analysis} to explore the symbolic representations
and generate the concrete values of variables,
which simultaneously make one data constraint hold and the other violated.
To prove the equivalence, we conduct the \emph{isomorphism analysis} with a tree isomorphism algorithm~\cite{AHU}
to examine whether the two symbolic representations
can be transformed into each other by reordering the clauses and commutative terms.
We combine the two analyses with the SMT solving,
which determines the logical equivalence of the symbolic representations,
finally obtaining a three-staged decision procedure.

We implement \ToolName\ and evaluate it upon a FinTech system in Company A,
which maintains \TotalRuleNumber\ data constraints in total.
Leveraging \ToolName,
we discover that \TotalEqRuleNumber\ data constraints have at least one equivalent variant in the system,
indicating that \TotalDeletedRuleNumber\ data constraints are redundant.
\ToolName\ finishes the equivalence clustering in three hours
and achieves the equivalence searching in 1.22 seconds per data constraint.
Except for the SMT solving, the stages of \ToolName\ can be proven to work in polynomial time.
We also prove the soundness and completeness of \ToolName\ theoretically for a given syntax of data constraints.
Our efforts have crossed the line from developing an academic-only decision procedure to one that is practical enough to be deployed.

In summary, we make the following major contributions:
\begin{itemize}[leftmargin=*]
	\item We formulate the data constraint equivalence problem,
	which is critical for the FinTech system maintenance.
	\item We propose a sound and complete decision procedure \ToolName\ to determine the data constraint equivalence,
	efficiently supporting the equivalence clustering and searching.
	\item We implement \ToolName\ and evaluate it upon the data constraints in a global FinTech company with 1 billion active users,
	showing that it efficiently detects a significant number of equivalent data constraints.
\end{itemize}
\section{Background and Motivation}
\label{subsec:usage}
This section presents the background and highlights the motivation of our work.

\begin{figure}[t]
	\centering
	\includegraphics[width=\linewidth]{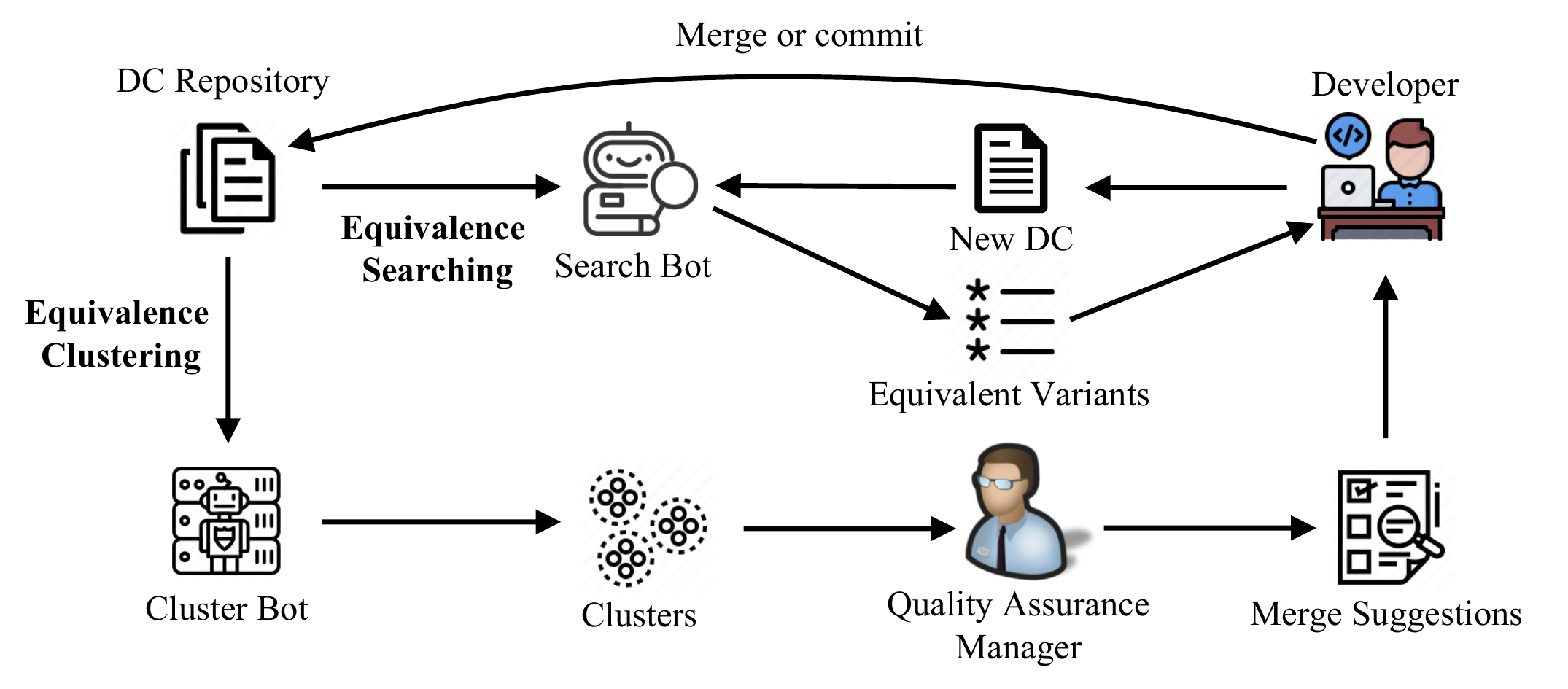}
	\caption{The workflow of equivalence searching and clustering}
	\label{fig:app}
\end{figure}

\begin{figure*}
	\centering
	\includegraphics[width=0.95\linewidth]{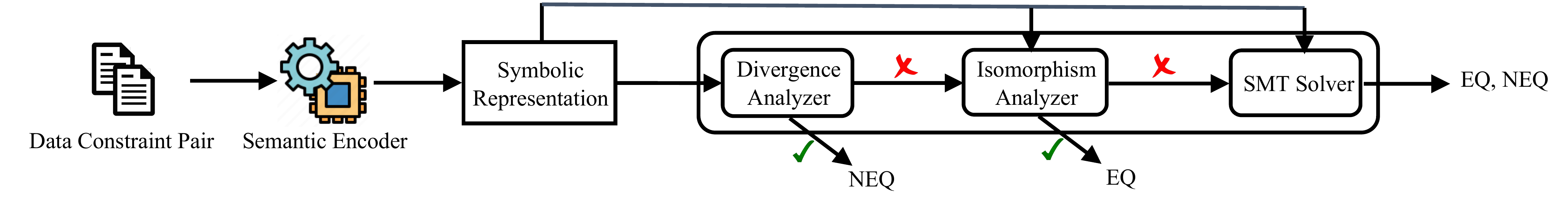}
	\caption{Schematic overview of our decision procedure \ToolName}
	\label{fig:workflow}
\end{figure*}

\subsection{Equivalent Data Constraints in FinTech Systems}
A FinTech system usually consists of a cluster of database-backed applications manipulating large amounts of data,
making the data records exhibit specific properties.
To improve the reliability, the developers often specify data constraints
to describe target properties
and set up a data reconciliation (DR) platform~\cite{bakhtouchi2020data} to examine them in the runtime.
Once a data constraint is violated, 
developers can receive detailed runtime information to guide further system diagnosis.

In reality, many development teams continuously submit data constraints to a central DR platform.
For example, around 100 teams in Company A actively submit data constraints daily to the platform.
Unaware of existing submissions,
developers often submit data constraints equivalent to existing ones.
Besides, the developers tend to be conservative about removing constraints,
as they do not want to risk missing data errors.
Finally, the accumulation of equivalent data constraints becomes the technical debt~\cite{ernst2021technical} of a FinTech system:
\begin{itemize}[leftmargin=*]
	\item The DR platform examines equivalent data constraints redundantly,
	which causes unnecessary resource consumption,
	e.g., CPU time, disk IO, and network traffic.
	\item Multiple alerts are fired to developers if equivalent data constraints are violated,
	which requires more time budget for diagnosis,
	lengthening the system's development cycle.
	\item All the equivalent data constraints should be updated if the table schema or the target properties are changed,
	involving extra labor in interacting with the DR platform.
\end{itemize}
Thus, it is crucial to tackle equivalent data constraints in the maintenance, 
which promotes resource-saving and eases the debugging and refactoring of a FinTech system.


\subsection{Resolving Equivalent Data Constraints}
To resolve the technical debt,
the developers of Company A propose two demands, namely equivalence clustering and equivalence searching,
to tackle equivalent data constraints.
Specifically, they expect to integrate two bots into the CI/CD workflow~\cite{ZampettiGBP21} of a FinTech system as follows.

\begin{itemize}[leftmargin=*]
	\item \emph{Equivalence searching}: 
	A developer commits a new data constraint to the bot for searching existing equivalent variants.
	The list of equivalent variants assists the developer in deciding whether to merge it with any existing one.
The workflow is shown in the upper part of Fig.~\ref{fig:app}.
	
	\item \emph{Equivalence clustering}: 
	A quality assurance (QA) manager exports all the data constraints to the bot, which
	divides the  data constraints into equivalence clusters.
	Then, the QA manager summarizes merge suggestions
	and sends them to developers for further confirmation.
	The lower part of Fig.~\ref{fig:app} shows the workflow of equivalence clustering.
\end{itemize}

Generally, the two bots resolve the redundancy from two perspectives, respectively.
First, the equivalence searching conducts the instant checking of newly-submitted data constraints, 
enabling the developers to avoid redundancy if possible.
Second, the equivalence clustering supports the nightly scan of the whole repository of data constraints.
The QA managers can inspect the clustering information to find opportunities for merging equivalent ones.
During the development cycle,
the two bots can serve as two lines of defense for redundancy issues in the CI/CD workflow.

To automate the overall workflow,
we need an efficient decision procedure 
to verify whether two data constraints are equivalent or not.
Specifically,
the two bots would invoke the decision procedure to determine the equivalence in the clustering and searching, 
respectively.
In this work, we aim to design an effective solution for verifying data constraint equivalence,
and promoting  data constraint maintenance with our decision procedure.

\section{\ToolName\ in a Nutshell}
\label{sec:overview}
This section presents a motivating example to show our insight (\cref{subsec:me})
and outlines our decision procedure (\cref{subsec:outline}).

\subsection{Motivating Examples}
\label{subsec:me}

Verifying the data constraint equivalence is non-trivial in industrial scenarios.
First, the cost of the decision procedure can accumulate significantly
due to the vast number of data constraints~\cite{YangSYC020}.
Second, the decision procedure can prune necessary data constraints
or miss equivalent ones
if it is not sound or complete,
increasing the risk of data security and hiding the opportunity for optimization.
Thus, we need to  simultaneously ensure the soundness, completeness, and efficiency of the decision procedure.

Fig.~\ref{fig:motivatingEx} shows four data constraints as examples.
Specifically, the data constraints in Fig.~\ref{subfig:motivatingEx1a} and Fig.~\ref{subfig:motivatingEx1b} depict the properties
where three attributes of the table $t$ have positive values in two cases,
and the values of \emph{oid} and \emph{iid} are not 0, respectively.
Besides, the data constraints in Fig.~\ref{subfig:motivatingEx1d} and Fig.~\ref{subfig:motivatingEx1c} describe the property
where the changes to the account balances are equal to the transferred cash amount,
and the ids of the two accounts, i.e., $iid$ and $oid$, are not 0.
According to the examples,
we can obtain the following two important observations:
\begin{itemize}[leftmargin=*]
	\item Non-equivalent data constraints tend to have different lexical tokens, such as database attributes, literals, and operators.
	For example, the data constraints in Fig.~\ref{subfig:motivatingEx1a} and Fig.~\ref{subfig:motivatingEx1b}
	examine different attributes.
	It is likely to generate the values of table attributes, making them evaluate differently.
	\item Equivalent data constraints often only differ in the orders of commutative operands
	and independent statements after eliminating user-defined variables. 
	For instance, the data constraints in Fig.~\ref{subfig:motivatingEx1d} and Fig.~\ref{subfig:motivatingEx1c}
	share the isomorphic syntactic structure,
	which implies their equivalence.
\end{itemize}
Based on the observations,
we realize that the lexical differences and syntactic isomorphism
enable us to efficiently refute and prove the equivalence, respectively.
If we generate ``good\text{''} concrete values making two data constraints evaluate differently
or find the isomorphism between the syntactic structures,
we can avoid SMT solving and achieve high efficiency.

\subsection{Outline of Decision Procedure}
\label{subsec:outline}
According to our insight,
we design an efficient, sound, and complete decision procedure for verifying  data constraint equivalence.
To depict the data constraint semantics,
we propose the semantic encoding to construct a FOL formula in a restrictive form as its symbolic representation,
which eliminates user-defined variables (e.g., the variable \emph{cash} in Fig.~\ref{subfig:motivatingEx1c}).
Based on the symbolic representations,
our decision procedure works in three stages, as shown in Fig.~\ref{fig:workflow}.
\begin{itemize}[leftmargin=*]
	\item The divergence analysis explores the symbolic representations with the guidance of lexical differences, aiming to generate concrete values that make data constraints evaluate differently.
	For example, it explores the clause induced by the last assertion in Fig.~\ref{subfig:motivatingEx1a}
	and assigns 0 to the attribute \emph{oid} to violate the assertion.
	Also, it concretizes the variables in Fig.~\ref{subfig:motivatingEx1b} to make the data constraint satisfied.
	\item The isomorphism analysis constructs the parse trees of the symbolic representations and examines whether the parse trees are isomorphic.
	The analysis abstracts away the order of commutative constructs, such as independent statements and commutative operands.
	For example, it discovers the isomorphic structures in Fig.~\ref{subfig:motivatingEx1d} and Fig.~\ref{subfig:motivatingEx1c}, blurring the orders of assertions and the operands of $+$ and $==$.
	\item If the first two analyses can not refute or prove the equivalence, we invoke an SMT solver to check the logical equivalence of the symbolic representations.
	To ensure soundness and completeness,
	we perform the SMT encoding with a decidable fragment in the combined theory of bit-vector, floating-point arithmetic, and string.
\end{itemize}

Apart from soundness and completeness,
\ToolName\ also features a theoretical guarantee of complexity.
The symbolic representation construction, the divergence analysis, and the isomorphism analysis
can work in polynomial time  to the size of the abstract syntax tree of a data constraint. 
Our evaluation also provides strong evidence of the \ToolName's high efficiency
in the equivalence clustering and searching.

\section{Problem Formulation}
\label{sec:preliminaries}
This section presents the syntax (\cref{subsec:syntax}) and formulates the data constraint equivalence problem (\cref{subsec:ps}).


\subsection{Data Constraint Syntax}
\label{subsec:syntax}

\begin{figure}[t]
	\[
	\begin{array}{lll}
		\mathcal{V}:= v_{d} \ | \ x \\
		\mathcal{L}:=\{l_i \ | \ i \geq 1 \}\\
		\mathcal{A}:=  l \ | \ v_d \ | \ a_1 \oplus a_2\\
		\mathcal{C}:= a_1 \odot a_2 \ | \ x_1 \odot x_2 \ | \ a \odot x \ | \ x \odot a \ | \ p(v, l) \ | \ p(v_1, v_2)\\
		\mathcal{B}:= c \ | \ b_1  \ \textbf{and} \ b_2 \ | \ b_1 \ \textbf{or} \ b_2 \ | \ \textbf{not} \ b \ | \ \textbf{ite}_b(c_0, b_1, b_2)\\
		\mathcal{S}:= x = a  \ | \ \textbf{assert}(b) \ | \ s_1; s_2 \ | \ \textbf{ite}_s(c_0, s_1, s_2)\\
		\mathcal{R}:=  s+\\
		\oplus:= \ + \ | \ - \ | \ \times \ | \ \div\\
		\odot:=  \ \ > \ | \ < \ | \ \geq \ | \ \leq \ | \ == \ | \ \neq\\
		\mathcal{P}:=  \{\text{prefixOf}, \ \text{suffixOf}, \ \text{contains}, \ \text{equals}\}\\
	\end{array}
	\]
	\vspace{-2mm}
	\caption{The syntax of data constraints}
	\label{lang-syntax}
	\vspace{-4mm}
\end{figure}

Fig.~\ref{lang-syntax} summarizes the  syntax.
A \emph{variable} is a data variable $v_d \in \mathcal{V}_d$ indicating the value of a table attribute,
or an user-defined variable $x \in \mathcal{V}_u$ storing the value temporally.
Its value can be a finite-length integer, a floating point number, or a string.
A \emph{literal} is a constant value.
An \emph{arithmetic expression} can be a literal, 
a data variable, 
or a compound arithmetic expression.
A \emph{comparison expression} compares arithmetic expressions and user-defined variables, or examines the strings with predicates $p \in \mathcal{P}$.
A \emph{Boolean expression} is a comparison expression
or a compound expression with logical connectives.
A \emph{statement} is an assignment, an assertion, a sequencing, or an $\textbf{ite}_{s}$ statement.
Particularly, the conditions in $\textbf{ite}_{b}$ expressions and $\textbf{ite}_{s}$ statements only relate to data variables.
Finally, a \emph{data constraint} consists of finite statements.
All its assertions are expected to hold for given database tables.

The syntax is expressive enough to specify target properties in real-world scenarios. It covers all the patterns in \cite{florez2022retrieving}, 
such as value comparison, conditional comparison, etc.
Also, user-defined variables support writing data constraints flexibly.
Arithmetic operations and string predicates support expressing complex properties, e.g., comparing the sums of cash amounts and matching between string variables.

\subsection{Data Constraint Equivalence Problem}
\label{subsec:ps}
Before stating the problem,
we first introduce the notions of interpretation and semantic equivalence as follows.

\begin{definition}
	\label{def:interpretation}
	An interpretation $I$ maps each data variable $v_{d}$ to a value in its domain.
	$I$ is a model of a data constraint $r$, denoted by $I \models r$,
	if all the assertions hold under $I$.
\end{definition}

\begin{example}
	\label{ex:interpretation}
	The following interpretation $I$ is a model of the data constraint in Fig.~\ref{subfig:motivatingEx1a}.
	\abovedisplayshortskip=2pt
\belowdisplayshortskip=2pt
\abovedisplayskip=2pt
\belowdisplayskip=2pt
	$$I = [t.ty \mapsto \text{`IN'}, t.in \mapsto 1, t.out \mapsto 0, t.oid \mapsto 1, t.amt \mapsto 1]$$
\end{example}

An interpretation indicates the values of table attributes.
A data constraint induces a set of interpretations making its assertions hold.
Formally, we define the semantic equivalence.

\begin{definition}
	\label{def:se}
	The data constraints $r_1$ and $r_2$ are semantically equivalent, denoted by $r_1 \simeq r_2$, if and only if
	\abovedisplayshortskip=2pt
\belowdisplayshortskip=2pt
\abovedisplayskip=2pt
\belowdisplayskip=2pt
	$$\forall I: I \models r_1 \Leftrightarrow I \models r_2$$
\end{definition}

\begin{example}
	Based on Example~\ref{ex:interpretation},
	we can construct 
	\abovedisplayshortskip=2pt
\belowdisplayshortskip=2pt
\abovedisplayskip=2pt
\belowdisplayskip=2pt
	$$I' = I[t.new \mapsto0, t.iid\mapsto 1]$$
	$I'$ is not a model of the data constraint in Fig.~\ref{subfig:motivatingEx1b},
	while it is a model of the data constraint in Fig.~\ref{subfig:motivatingEx1a},
	indicating that they are not semantically equivalent.
\end{example}

In this work,
we aim to propose a decision procedure to verify whether $r_1$ is semantically equivalent to $r_2$
for a given data constraint pair $(r_1, r_2)$.
However, finding a sound,  complete, and efficient solution is challenging. Theoretically, any instance of SAT problem~\cite{Cook71} can be reduced to an instance of our problem by constructing two proper data constraints in polynomial time. Formally, we state the complexity barrier of our problem as follows.

\begin{theorem}
	\label{thm:decomp}
	Data constraint equivalence problem is NP-hard.
\end{theorem}

\textbf{Roadmap.}
To verify the equivalence,
we propose a symbolic representation to encode the semantics~(\cref{sec:as}) and design an efficient decision procedure (\cref{sec:dp}).
Particularly, we introduce light-weighted reasoning to refute and prove the equivalence efficiently,
which is our main technical contribution.
By fusing our light-weighted reasoning with SMT-based analysis,
our decision procedure features soundness and completeness, and  achieves high efficiency in supporting the equivalence clustering and searching.
\section{Semantic Encoding}
\label{sec:as}
This section introduces the symbolic representation to depict the semantics (\cref{subsec:sr}),
presents the symbolic evaluation (\cref{subsec:evalnorm}),
and summarizes the benefit at the end (\cref{subsec:sesummary}).

\subsection{Symbolic Representation}
\label{subsec:sr}
A data constraint is essentially a program with data variables as inputs.
The values of data variables determine the values of all the variables and expressions.
Based on the intuition, we propose the concepts of symbolic terms and conditions
to depict the values of variables and expressions.

\begin{definition}
	A symbolic term $\tau$ represents the value of a variable or a literal
	in either of the forms:
	\begin{itemize}[leftmargin=*]
		\item $\tau := v_d$ or $\tau := l$ is a data variable or a literal, respectively.
		\item $\tau := \tau_1 \oplus \tau_2$ is a compound term with an arithmetic operator.
	\end{itemize}
\end{definition}

\begin{definition}
	A symbolic condition $\phi$ is a FOL formula in one of the following forms:
	\begin{itemize}[leftmargin=*]
		\item An atomic condition is an arithmetic comparison of two symbolic terms
		or a string comparison,
		i.e., $\phi := \tau_1 \odot \tau_2$ or $\phi := p(\tau_1, \tau_2)$, where $p \in \mathcal{P}$ is a string predicate.
		\item A compound condition is a FOL formula with logical connectives,
		i.e., $\phi := \phi_1 \land \phi_2$, $\phi := \phi_1 \lor \phi_2$, or $\phi := \lnot \phi_0$.
	\end{itemize}
\end{definition}

\begin{example}
	In Fig.~\ref{subfig:motivatingEx1c},
	the values of \emph{cash} can be represented by the terms $t.out + t.new$ and $t.new - t.in$.
	The condition of the $\textbf{ite}_s$ statement is encoded by $\lnot \texttt{contains}(t.ty, \ \text{`IN'})$.
\end{example}

The symbolic terms represent the values of variables, literals, and arithmetic expressions,
while the symbolic conditions encode the values of Boolean expressions,
providing the ingredient for defining the symbolic representations.

\begin{definition}
	\label{def:sr}
	For a data constraint $r$,
	its symbolic representation is a symbolic condition $\varphi$ satisfying
	\begin{itemize}[leftmargin=*]
		\item For any interpretation $I$, $I \models r$ if and only if $I \models \varphi$. 
		\item The negations only occur before string atomic constraints. 
	\end{itemize}
\end{definition}

Intuitively, the symbolic representation encodes the semantics faithfully with a FOL formula,
which only relates to data variables and exclude redundant negations.
It abstracts away user-defined variables and blurs syntactic differences in terms of negations effectively,
enabling us to design light-weighed reasoning for equivalence verification.
In what follows, we show how to construct the symbolic representation in detail.

\subsection{Symbolic Evaluation}
\label{subsec:evalnorm}
Now we propose the symbolic evaluation to construct the symbolic representation.
Basically, the symbolic evaluation consists of two stages,
which collects the values of Boolean expressions in each assertion,
and eliminates unnecessary negations, respectively.
Before delving into details,
we first introduce the notion of the symbolic state.

\begin{definition}
	\label{def:symstate}
	Given a data constraint $r$,
	the symbolic state $\mathbf{S}$ at program location $\ell$ is $(\mathbf{E}, \mathbf{\Phi})$,
	where 
	\begin{itemize}[leftmargin=*]
		\item An \emph{environment} $\mathbf{E}$ maps a variable $v$ or an arithmetic expression $e$ to a term-condition pair set $\{(\tau, \phi)\}$,
		indicating that $v$ or $e$ evaluates to the same value of $\tau$ when $\phi$ holds.
		\item A \emph{property} $\mathbf{\Phi}$ is a symbolic condition that summarizes the assertions in $r$ before the program location $\ell$.
	\end{itemize}
\end{definition}

\begin{example}
	After the first assertion in Fig.~\ref{subfig:motivatingEx1c},
	we have 
	\abovedisplayshortskip=2pt
\belowdisplayshortskip=2pt
\abovedisplayskip=2pt
\belowdisplayskip=2pt
	$$\mathbf{E} = [t.iid \mapsto \{(t.iid, T)\}, \ 0 \mapsto \{(0, T)\}]\ \ \mathbf{\Phi} = (t.iid \neq 0)$$
\end{example}

Now we present the technical details of the symbolic evaluation.
In the first stage,
we evaluate the variables and expressions to obtain a FOL formula depicting the semantics,
which only relates to the data variables. 
Specifically,
we define the evaluation rules in Fig.~\ref{rule:statement} and Fig.~\ref{rule:boolexpr}.

\begin{itemize}[leftmargin=*]
	\item The rule \texttt{ASSIGN} evaluates the RHS with the rules \texttt{VAR} and \texttt{AE} in Fig.~\ref{rule:boolexpr},
	and applies the strong update to $\mathbf{E}$, enforcing the user-defined variable $v$ and the expression $a$ have the same value.
	It successfully evaluates user-defined variables, making the symbolic terms only relate to the data variables. 
	\item The rule \texttt{ASSERT} evaluates the Boolean expression $b$ to a symbolic condition $\psi$.
	It then connects $\psi$ and the original property $\mathbf{\Phi}$ with a logical conjunction.
	This, in turn, forms a property that accumulates the conditions of the assertions.  
	\item The rules \texttt{SEQ} and \texttt{ITE-S} are defined straightforwardly.
	\texttt{SEQ} applies the evaluation rules of two components sequentially.
	\texttt{ITE-S} evaluates the two cases separately and joins two symbolic states according to the branch condition.  
\end{itemize}

\begin{figure}[t]
	\[
	\inference[\texttt{ASSIGN}]
	{
		\mathbf{E} \vdash_e a \leadsto V \ \ \ \mathbf{E}' = \mathbf{E}[v \mapsto V]
	}
	{
		\mathbf{E}, \mathbf{\Phi} \vdash v = a  \leadsto \mathbf{E}', \mathbf{\Phi}
	}
	\]
	\[
	\inference[\texttt{ASSERT}]
	{
		\mathbf{E} \vdash_b b \leadsto \psi \ \ \ \mathbf{\Phi}' = \mathbf{\Phi} \land \psi
	}
	{
		\mathbf{E}, \mathbf{\Phi} \vdash \textbf{assert}(b)  \leadsto \mathbf{E}, \mathbf{\Phi}'
	}
	\]
	\[
	\inference[\texttt{SEQ}]
	{
		\mathbf{S} \vdash s_1 \leadsto \mathbf{S}_1 \ \ \  \mathbf{S}_1 \vdash s_2 \leadsto \mathbf{S}'\\
	}
	{
	\mathbf{S} \vdash s_1; s_2  \leadsto \mathbf{S}'
	}
	\]
	\[
	\inference[\texttt{ITE-S}]
	{
		\mathbf{E} \vdash_b c_0 \leadsto \gamma_1  \ \ \gamma_2 = \lnot \gamma_1 \ \ \mathbf{E},\mathbf{\Phi} \vdash s_i \leadsto \mathbf{E}_i, \mathbf{\Phi}_i\\
		\mathbf{E}' = [u \mapsto \bigcup_{i = 1}^{2}\{(\tau_i, \phi_i \land \gamma_i) | (\tau_i, \phi_i) \in \mathbf{E}_i(u)\}]\\
	}
	{
		\mathbf{E}, \mathbf{\Phi} \vdash \textbf{ite}_s(c_0, s_1, s_2)  \leadsto \mathbf{E}',  ite(\gamma_1, \mathbf{\Phi}_1, \mathbf{\Phi}_2)
	}
	\]
	\vspace{-2mm}
	\caption{Evaluation rules of statements}	
	\label{rule:statement}
\end{figure}
\begin{figure}[t]	
	\[
\inference[\texttt{VAR}]
{
	u \in \mathcal{L} \cup \mathcal{V}_d \ \ \ U = \{(u, T)\}\\
}
{
	\mathbf{E} \vdash_e u\leadsto U
}
\]
\[
\inference[\texttt{AE}]
{
	a_i \in \mathcal{A} \ \ \ \mathbf{E} \vdash_e a_i \leadsto U_i\\
	A = \{(t_1 \oplus t_2, \phi_1 \land \phi_2) \ | \ (t_i, \phi_i) \in U_i\}\\
}
{
	\mathbf{E} \vdash_e a_1 \oplus a_2 \leadsto A
}
\]
\[
\inference[\texttt{ACmp}]
{
	u_i \in \mathcal{A} \cup \mathcal{V}_u \ \ \ 
	\mathbf{E} \vdash_e u_i \leadsto U_i\\
	B = \{(t_1 \odot t_ 2) \land \phi_1 \land \phi_2 \ | \ (t_i, \phi_i) \in U_i\}
}
{
	\mathbf{E} \vdash_b u_1 \odot u_2 \leadsto \bigvee_{\phi \in B} \phi
}
\]
\[
\inference[\texttt{ITE-E}]
{
	\mathbf{E} \vdash_b c_0 \leadsto \gamma_0\ \ \ 
	\mathbf{E}  \vdash_b b_i \leadsto \gamma_i\\
}
{
	\mathbf{E} \vdash_b \textbf{ite}_b(c_0, b_1, b_2) \leadsto  (\gamma_1 \land \gamma_0) \lor (\gamma_2 \land \lnot \gamma_0)\\
}
\]
	\vspace{-2mm}
	\caption{Helper rules evaluating expressions}
	\label{rule:boolexpr}
	\vspace{-3mm}
\end{figure}

We omit the rules of evaluating string comparisons and other compound Boolean expressions due to limited space,
which are similar to the rules \texttt{ACmp} and \texttt{ITE-E}.
Based on the rules,
we evaluate a data constraint stepwise.
Initially, the symbolic state
is a pair of empty mapping and a \emph{true} value.
By applying the rule of each statement along control flow paths,
we obtain the symbolic state at each program location
and finally summarize all the assertions with the property $\mathbf{\Phi}_e$ at the exit,
which depicts the semantics of the data constraint.


\begin{example}
	\label{ex:er}
	Consider the data constraint in Fig.~\ref{subfig:motivatingEx1c}.
	We obtain $\mathbf{\Phi} = \phi_1 \land ((\phi_2 \land \phi_4)\lor (\phi_3 \land \lnot \phi_4))$ at its exit, where
	\abovedisplayshortskip=0pt
	\belowdisplayshortskip=0pt
	\abovedisplayskip=2pt
	\belowdisplayskip=2pt
	\begin{align*}
		\phi_1&=(t.iid \neq 0) \land (t.oid \neq 0) \ \phi_2=(t.out + t.new = t.old)\\
		\phi_3&=(t.new - t.in = t.old) \ \ \phi_4 = \lnot \texttt{contains}(t.ty, \  \text{`IN'})
	\end{align*}
\end{example}

In the second stage, we eliminate the negations in $\mathbf{\Phi}_e$ that do not apply to atomic string constraints.
Technically, we first transform $\mathbf{\Phi}_e$ into the negation normal form (NNF),
in which the negation applies only to atomic formulas.
Then, we eliminate the negation before each atomic arithmetic constraint by changing the comparison operator,
e.g., transforming $\lnot (t.a \geq t.b)$ to $t.a < t.b$. 
Notably, the above transformations can be achieved by the breadth-first search upon the parse tree of $\mathbf{\Phi}_e$,
where the symbolic representation is constructed on the fly.
The overall time complexity is linear to the size of $\mathbf{\Phi}_e$.

\begin{example}
	In Example~\ref{ex:er}, we eliminate the negations and get
	the symbolic representation $\varphi = \phi_1 \land((\phi_2 \land \phi_4) \lor \phi')$, where
	$\phi'=(t.new - t.in = t.old) \land \texttt{contains}(t.ty, \  \text{`IN'})$.
\end{example}


\subsection{Summary}
\label{subsec:sesummary}
The symbolic representation is essentially a Boolean function of data variables,
featuring the following three benefits:
\begin{itemize}[leftmargin=*]
	\item The symbolic representations preserve the lexical differences in terms of data variables, literals, and operators,
	which can indicate the possible non-equivalence.
	\item The symbolic evaluation evaluates the user-defined variables,
	abstracting away the difference in terms of their names, which do not affect the semantics.
	\item The elimination of unnecessary negations normalizes the FOL formulas 
	and yields isomorphic symbolic representations for more equivalent data constraints.
\end{itemize}
Thus, the semantic encoding exposes lexical differences and syntactic isomorphism for light-weight reasoning,
which efficiently refutes and proves the equivalence (\cref{subsec:da}~\cref{subsec:ia}).
\section{Decision Procedure}
\label{sec:dp}
In this section, we first introduce the divergence analysis (\cref{subsec:da}) and isomorphism analysis (\cref{subsec:ia}) for efficiently refuting and proving the equivalence, respectively.
We then combine the two analyses with SMT solving to establish the decision procedure (\cref{subsec:ev}).
In what follows, we denote the data constraints by $r_1$ and $r_2$ and their symbolic representations by $\varphi_1$ and $\varphi_2$ for demonstration.

\subsection{Divergence Analysis}
\label{subsec:da}
Based on Definition~\ref{def:sr},
$\varphi_1$ and $\varphi_2$ depict the semantics of two data constraints faithfully.
We can safely refute the equivalence if there exists an interpretation $I$ making them evaluate to different truth values.
However, it is non-trivial to obtain such a desired interpretation efficiently.
The random sampling may hit a desired interpretation successfully after failing many attempts,
which can degrade the efficiency significantly.
To resolve the problem,
we attempt to explore specific Boolean structures of $\varphi_1$ and $\varphi_2$ and concretize the data variables within the structures.
Formally, we introduce the \textit{degrees of freedom} to guide the exploration.

\begin{definition}
	\label{def:freedom}
	For two symbolic representations $\varphi_1$ and $\varphi_2$, 
	the degrees of freedom of a clause $\phi$ occurring in $\varphi_1$ is
	\abovedisplayshortskip=0pt
	\belowdisplayshortskip=0pt
	\abovedisplayskip=2pt
	\belowdisplayskip=2pt
	$$\mathcal{DF}(\phi \ | \ \varphi_1, \varphi_2) = \frac{1}{h(\phi)} \cdot \sum_{M \in \{ V_d, L, O \}}|M(\phi) \setminus M(\varphi_2)|$$
	$h(\psi)$ is the height of the parse tree of $\psi$.
	$V_d(\psi)$ contains the data variables in $\psi$ but excludes arithmetic operands.
	$L(\psi)$ and $O(\psi)$ contain the literals and operators in $\psi$, respectively.
\end{definition}

Intuitively, 
a larger degrees of freedom indicates a higher possibility of making $\phi$ evaluate to a target truth value:
\begin{itemize}[leftmargin=*]
	\item First, a smaller value of $h(\phi)$ indicates the opportunity of finding the desired interpretation with fewer explorations.
	\item Second, a larger value of $|M(\phi) \setminus M(\varphi_2)|$ indicates that 
	$\phi$ has more unique lexical tokens absent in $\varphi_2$.
	The concretizations of data variables in $\phi$ and $\varphi_2$ are less intertwined.
	\item Third, $V_d(\phi)$ excludes arithmetic operands,
	as arithmetic operations can increase the difficulty of concretization.
\end{itemize}

\begin{example}
	\label{ex:s}
	Consider the data constraints in Fig.~\ref{subfig:motivatingEx1a} and Fig.~\ref{subfig:motivatingEx1b}.
	According to~\cref{sec:as}, their symbolic representations are
	\abovedisplayshortskip=0pt
	\belowdisplayshortskip=0pt
	\abovedisplayskip=2pt
	\belowdisplayskip=2pt
	\begin{align*}
		\varphi_1 &= ((t.in > 0 \land \phi_c) \lor (t.out > 0 \land \lnot \phi_c))\land \phi_a \land \phi_o\\
		\varphi_2 &= ((t.out > 0 \land \lnot \phi_c) \lor (t.new > 0 \land \phi_c))\land \phi_a \land \phi_i
	\end{align*}
	where $\phi_a = (t.amt >0)$,
	$\phi_o = (t.oid \neq 0)$, $\phi_i = (t.iid \neq 0)$,
	and $\phi_c = \texttt{contains}(t.ty, \  \text{`IN'})$.
	Let $\phi$ denote the first clause of $\varphi_1$.
	We have $V_d(\phi) \setminus V_d(\varphi_2) = \{t.in\}$,
	$L(\phi) \subseteq L(\varphi_2)$, and $O(\phi) \subseteq O(\varphi_2)$
	Thus, we have $\mathcal{DF}(\phi \ | \ \varphi_1, \varphi_2) = \frac{1}{3}$.
	Similarly, we have
	$\mathcal{DF}(\phi_a \ | \ \varphi_1, \varphi_2) = 0$ and $\mathcal{DF}(\phi_o \ | \ \varphi_1, \varphi_2) = 1$.
\end{example}

\setlength{\textfloatsep}{0.3cm}
\setlength{\floatsep}{0.3cm}
\begin{algorithm}[t]
	\SetNoFillComment
	\caption{Divergence analysis}
	\label{alg:da}
	\KwIn{$\varphi_1, \varphi_2$: Two symbolic representations\;}
	\KwOut{Whether $\exists I: \lnot(I \models \varphi_1 \leftrightarrow I \models \varphi_2)$}
	\SetKwFunction{generateAST}{generateAST}
	\SetKwFunction{serializeByBFS}{serializeByBFS}
	\SetKwFunction{AExpr}{A-Expr}
	\SetKwFunction{Predicate}{Predicate}
	\SetKwFunction{Statement}{Statement}
	\SetKwFunction{applyAE}{applyAERule}
	\SetKwFunction{applyP}{applyPRule}
	\SetKwFunction{applyS}{applySRule}
	\SetKwFunction{genOrbit}{PIE}
	\SetKwFunction{constructAlgRep}{constructAlgReps}
	\SetKwFunction{constructSymRep}{constructSymReps}
	\SetKwFunction{getDBVariables}{$\pi_v$}
	\SetKwFunction{getDBFunctions}{$\pi_f$}
	\SetKwFunction{SMTSolve}{SMT-Solve}	
	\SetKwFunction{enforce}{\textit{explore}}	
	\SetKwFunction{enforceTitle}{\textbf{explore}}
	\SetKwFunction{collectDVar}{collectVar}
	\SetKwFunction{Vars}{Var}
	\SetKwFunction{findClause}{findClause}
	\SetKwFunction{collectFreeDVar}{FreeVar}
	\SetKwFunction{genValue}{concretize}
	\SetKwFunction{checkTruthValue}{check}
	\SetKwProg{myproc}{Procedure}{}{}	
	\ForEach{$(\phi_1, \phi_2) \in \{(\varphi_1, \varphi_2), (\varphi_2, \varphi_1)\}$}{
		$I \leftarrow \bot$\;
		$status \leftarrow T$\;
		\enforce{$\phi_1$, $\phi_1$, $\phi_2$, $F$}\;
		\enforce{$\phi_2$, $\phi_2$, $\phi_1$, $T$}\;
		\uIf{$status$ is $T$}{
			\Return{\textit{true}}\;
		}
	}
	\Return{\textit{unknown}}\;
	\myproc{\enforceTitle{$\phi$, $\varphi$, $\varphi'$, $tv$}}{
		\If{$status$}{
			\uIf{$\phi$ is atomic}{
				\uIf{\collectFreeDVar{$\phi$, $I$} $\neq \emptyset$}{
					$I \leftarrow$ \genValue{$\phi, tv$}\;
				}\Else{$status \leftarrow$ \checkTruthValue{$I \models \phi = tv$}\;}
			}
		   	\uElseIf{$(LC(\phi), tv) \in \{(\land, T), (\lor, F)\}$}{
			   	\ForEach{$\phi_i \in C(\phi)$}{
			   		\enforce{$\phi_i$, $\varphi$, $\varphi'$, $tv$}\;
			   	}
			}\uElse{
				$\phi' \leftarrow \mathop{\arg\max}_{\phi_i \in C(\phi)} \ \ \mathcal{DF}(\phi_i \ | \ \varphi, \varphi')$\;
				\enforce{$\phi'$, $\varphi$, $\varphi'$, $tv$}\;
			}
		}
	}
\end{algorithm}

Based on the degrees of freedom,
we propose the divergence analysis to generate a desired interpretation.
Alg.~\ref{alg:da} shows its technical details.
It receives two symbolic representations $\varphi_1$ and $\varphi_2$
and attempts to generate a desired interpretation enforcing them evaluate differently (lines 1--7).
The function \textit{explore} traverses the clauses level by level (lines 9--21),
handling three kinds of clauses $\phi$ with specific strategies:
\begin{itemize}[leftmargin=*]
	\item If $\phi$ is atomic,
	we concretize the free variables to make $\phi$ evaluate to $tv$ (lines 12--13).
	If there is no free variable, we check whether $\phi$ evaluates to $tv$ under $I$ (line 15).
	\item If $\phi$ is a connected with $\land$ and $tv$ is \emph{true}, or $\phi$ is connected with $\lor$ and $tv$ is \emph{false},
	we explore all the clauses in $\phi$ and enforce them evaluates to $tv$ (lines 16--18).
	\item Otherwise, we select the clause $\phi'$ with the maximal degrees of freedom
	and enforce it evaluate to $tv$ (lines 20--21).
\end{itemize}
If each clause evaluates to the target value,
Alg.~\ref{alg:da} finds the desired interpretation, thereby refuting the equivalence.

\begin{example}
\label{ex:sr}
In Example~\ref{ex:s}, $\varphi_1$ is connected with the logical conjunction.
We only need to select and explore one of its clauses if we want to make $\varphi_1$ evaluate to \textit{false}.
The third clause $\phi_o$ has a larger degrees of freedom than the other two,
so we select it and assign 0 to $t.oid$.
Similarly, we can enforce $\varphi_2$ evaluate to \emph{true},
which finally refutes the equivalence.
\end{example}

Lastly, it is worth mentioning that the data constraints with different lexical tokens
are often non-equivalent,
while it is unsound to refute the equivalence directly based on lexical differences.
In contrast, our divergence analysis essentially utilizes lexical differences to guide the interpretation generation,
which supports refuting the equivalence soundly.

\subsection{Isomorphism Analysis}
\label{subsec:ia}

As the FOL formulas,
the symbolic representations $\varphi_1$ and $\varphi_2$ are logically equivalent
if we can transform $\varphi_1$ to $\varphi_2$ by reordering commutative sub-formulas and terms in $\varphi_1$.
For example, the evaluation of a FOL formula does not depend on
the order of the clauses connected with the logical disjunction and conjunction.
Also, any permutation of the operands of commutative arithmetic operators, such as addition and multiplication,
always yields the logically equivalent formula.
In other words, we can prove the data constraint equivalence safely
by identifying the isomorphism between $\varphi_1$ and $\varphi_2$.

Based on the above key idea,
we propose the isomorphism analysis to determine whether the parse trees of $\varphi_1$ and $\varphi_2$ are isomorphic,
which is formulated in Alg.~\ref{alg:ia}.
Using the AHU algorithm~\cite{AHU} for tree isomorphism checking,
Alg.~\ref{alg:ia} proves the data constraint equivalence if the parse trees are isomorphic (lines 1--2).
Particularly, the functions \textit{SCT} and \textit{STT} process the clauses and terms of a symbolic representation
in a top-down manner, respectively,
creating tree nodes and leaf nodes in the parse tree.

\begin{algorithm}[t]
	\SetNoFillComment
	\caption{Isomorphism analysis}
	\label{alg:ia}
	\KwIn{$\varphi_1, \varphi_2$: Two symbolic representations\;}
	\KwOut{Whether $\forall I: I \models \varphi_1 \leftrightarrow I \models \varphi_2$}
	\SetKwFunction{generateAST}{generateAST}
	\SetKwFunction{serializeByBFS}{serializeByBFS}
	\SetKwFunction{AExpr}{A-Expr}
	\SetKwFunction{Predicate}{Predicate}
	\SetKwFunction{Statement}{Statement}
	\SetKwFunction{applyAE}{applyAERule}
	\SetKwFunction{applyP}{applyPRule}
	\SetKwFunction{applyS}{applySRule}
	\SetKwFunction{genOrbit}{PIE}
	\SetKwFunction{constructAlgRep}{constructAlgReps}
	\SetKwFunction{constructSymRep}{constructSymReps}
	\SetKwFunction{getDBVariables}{$\pi_v$}
	\SetKwFunction{getDBFunctions}{$\pi_f$}
	\SetKwFunction{SMTSolve}{SMT-Solve}	
	\SetKwFunction{enforce}{enforce}	
	\SetKwFunction{enforceTitle}{\textbf{enforce}}
	\SetKwFunction{constructTree}{\textit{SCT}}	
	\SetKwFunction{constructTreeTitle}{\textbf{SCT}}
	\SetKwFunction{checkTreeIsomorphism}{checkTreeIsomorphism}
	\SetKwFunction{collectDVar}{collectDVar}
	\SetKwFunction{DVars}{DVars}
	\SetKwFunction{emptyTree}{emptyTree}
	\SetKwFunction{setTreeNode}{setTreeNode}
	\SetKwFunction{addChild}{addChild}
	\SetKwFunction{addChildren}{addChildren}
	\SetKwFunction{addLeaf}{Leaf}
	\SetKwFunction{addLeaves}{addLeaves}
	\SetKwFunction{findClause}{findClause}
	\SetKwFunction{collectFreeDVar}{getFreeDVar}
	\SetKwFunction{genValue}{genValue}
	\SetKwFunction{parseTerm}{\textit{STT}}
	\SetKwFunction{AHUcheck}{AHUcheck}
	\SetKwFunction{getAllOps}{Op}
	\SetKwFunction{flip}{flip}
	\SetKwFunction{getAtomicTerms}{Operand}
	\SetKwFunction{parseTermTitle}{\textbf{STT}}
	\SetKwFunction{checkTruthValue}{checkTruthValue}
	\SetKwFunction{Tree}{Tree}
	\SetKwProg{myproc}{Procedure}{}{}	
	\uIf{\AHUcheck{$\constructTree(\varphi_1)$, $\constructTree(\varphi_2)$}}{
		\Return{true}\;
	}
	\Return{unknown}\;
	\myproc{\constructTreeTitle{$\phi$}}{
		\uIf{$\phi: \phi_1 \circledast \cdots \circledast \phi_k$ \textbf{and} $\circledast \in \{\land, \lor, \lnot\}$}{
			$\Return \ \Tree(\circledast,  \{\constructTree(\phi_i) \ | \ 1 \leq i \leq k \})$\;
		}\uElseIf{$\phi: \tau_1 \circledast \tau_2$ \textbf{or} $\phi: \circledast(\tau_1, \tau_2)$}{
			\uIf{$\circledast \in \{ =, \neq, equals \}$}{
				$\Return \ \Tree(\circledast,  \{ \parseTerm(\tau_1),  \parseTerm(\tau_2)\})$;
			}\uElseIf{$\circledast \in \{<, \leq\}$}{
			    $\circledast' \leftarrow \flip(\circledast)$\;
				$\Return \ \addLeaf(\circledast',  \parseTerm(\tau_1),  \parseTerm(\tau_2))$;
			} \uElse{$\Return \ \addLeaf(\circledast,  \parseTerm(\tau_1),  \parseTerm(\tau_2))$;}
		}
	}
	\myproc{\parseTermTitle{$\tau$}}{
		\uIf{$\getAllOps(\tau) = \{\circledast\}$ \ \textbf{and} \ $\circledast \in \{+, *\}$}{
			$\Return\ \Tree(\circledast,  \getAtomicTerms(\tau))$\;
		}\uElseIf{$\tau: \tau_1 \oplus \tau_2$}{
			$\Return\ \addLeaf(\oplus, \parseTerm(\tau_1), \parseTerm(\tau_2))$\;
		}\Else{$\Return\ \addLeaf(\tau)$\;}
	}
\end{algorithm}

\begin{itemize}[leftmargin=*]
	\item When processing a non-atomic formula $\varphi$,
	\textit{SCT} creates a tree node to store the logical connective,
	and appends all the parse trees of its clauses (lines 5--6).
	\item For an atomic condition,
	\textit{SCT} creates a tree node 
	if the comparison operator is in $\{=, \neq\}$ or the string predicate is \emph{equals} (lines 8--9).
	Otherwise, it adds a leaf node to make sub-trees nonexchangeable (lines 10--14).
	Notably, it normalizes inequalities to enforce them using $>$ and $\geq$ only,
	which supports discovering more equivalent inequalities.
	\item 
	\textit{STT} constructs a tree node if a term $\tau$ only uses addition or multiplication (lines 16--17).
	For other cases, \textit{STT} creates a leaf node (lines 18--21).
\end{itemize}

\begin{figure}[h]
	\vspace{-5mm}
	\centering
	\subcaptionbox{\label{subfig:tree1}}{\includegraphics[width = .48\linewidth]{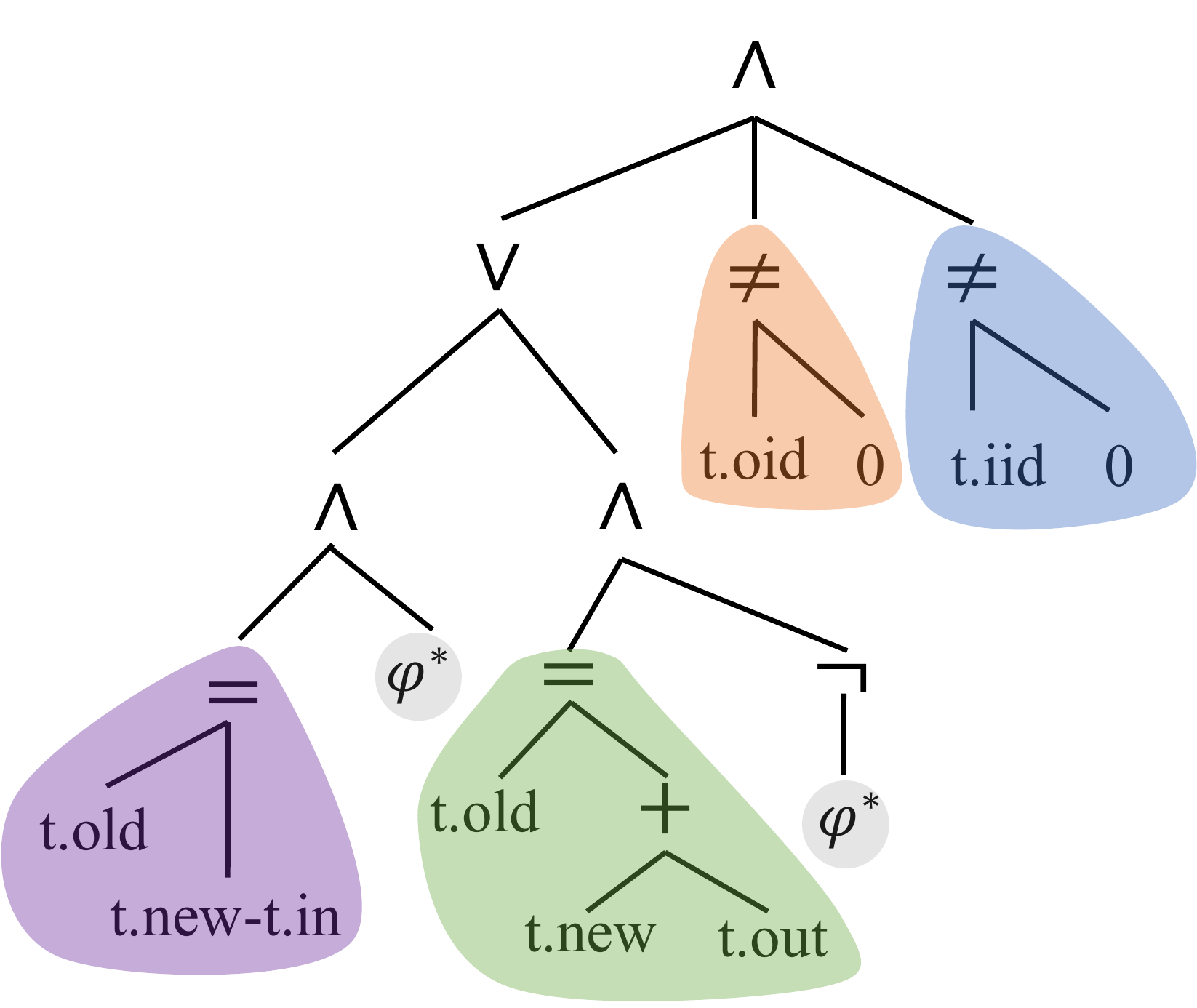}}
	\subcaptionbox{\label{subfig:tree2}}{\includegraphics[width = .48\linewidth]{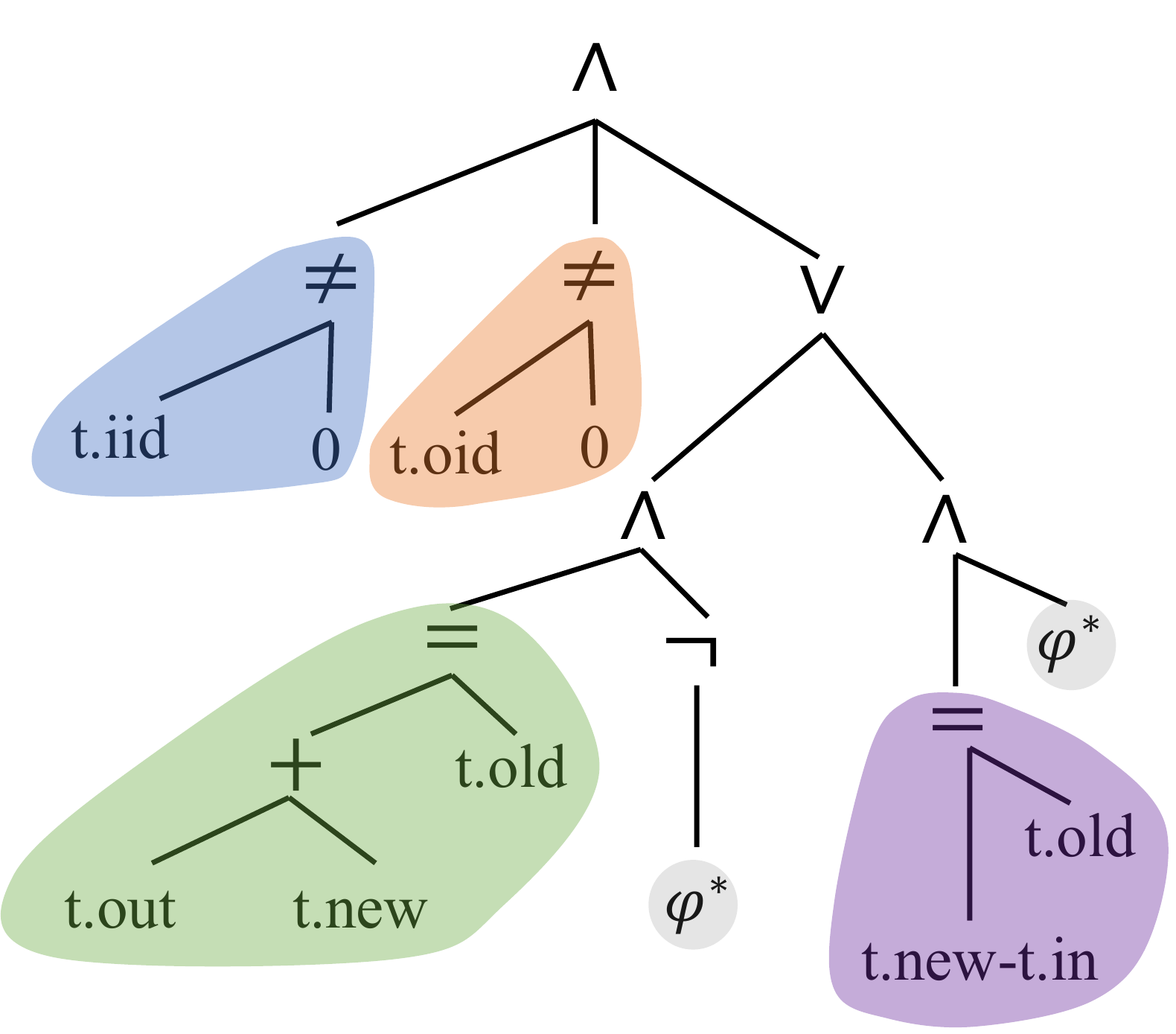}}
	\caption{Two isomorphic parse trees}
	\label{fig:tree}
\end{figure}

\begin{example}
	Fig.~\ref{subfig:tree1} and Fig.~\ref{subfig:tree2} show the parse trees of the symbolic representations for the data constraints 
	in Fig.~\ref{subfig:motivatingEx1d} and Fig.~\ref{subfig:motivatingEx1c}, respectively. 
	$\varphi^{*}$ represents $\texttt{contains}(t.ty, \  \text{`IN'})$.
	Their isomorphism proves the data constraint equivalence.
\end{example}

It is worth noting that the AHU algorithm in Alg.~\ref{alg:ia} is slightly different
from the standard one~\cite{AHU}.
Originally, the AHU algorithm sorts the sub-trees by level, as the orders of the sub-trees do not matter.
In our case, however, only the sub-trees of tree nodes can be arbitrarily permuted.
Thus, we modify the AHU algorithm to adapt it to the isomorphism analysis, not sorting the sub-trees of each leaf node.

\vspace{-1mm}
\subsection{Equivalence Verification with \ToolName}
\label{subsec:ev}

\begin{algorithm}[t]
	\SetNoFillComment
	\caption{Decision procedure}
	\label{alg:eqr}
	\KwIn{$r_1, r_2$: Two data constraints\;}
	\KwOut{Whether $r_1 \simeq r_2$ or not}
	\SetKwFunction{generateAST}{generateAST}
	\SetKwFunction{serializeByBFS}{serializeByBFS}
	\SetKwFunction{AExpr}{A-Expr}
	\SetKwFunction{Predicate}{Predicate}
	\SetKwFunction{Statement}{Statement}
	\SetKwFunction{applyAE}{applyAERule}
	\SetKwFunction{applyP}{applyPRule}
	\SetKwFunction{applyS}{applySRule}
	\SetKwFunction{genOrbit}{PIE}
	\SetKwFunction{constructSymRep}{getSymReps}
	\SetKwFunction{getDBVariables}{$\pi_v$}
	\SetKwFunction{getDBFunctions}{$\pi_f$}
	\SetKwFunction{SMTSolve}{SMT-Solve}	
	\SetKwFunction{isDivergent}{\textit{Divergent}}
	\SetKwFunction{isIsomorphic}{\textit{Isomorphic}}		
	$\varphi_1, \varphi_2 \leftarrow  \constructSymRep(r_1, r_2)$\;
	\If{\isDivergent{$\varphi_1, \varphi_2$} is true}{
		\Return{false}\;
	}
	\If{\isIsomorphic{$\varphi_1, \varphi_2$} is true}{
		\Return{true}\;
	}
	\Return{(\SMTSolve{$\lnot (\varphi_1 \leftrightarrow \varphi_2)$} is \textit{UNSAT})}\;
\end{algorithm}

Combining the above analyses with the SMT solving,
we obtain the decision procedure \ToolName\ in Alg.~\ref{alg:eqr}.
We first construct the symbolic representations $\varphi_1$ and $\varphi_2$ via the semantic encoding.
If we can not refute or prove the equivalence with the first two analyses (lines 2--5),
an SMT solver examines whether $\varphi_1$ and $\varphi_2$ are logically equivalent (line 6) for general cases.
Notably, the function \textit{Divergent} invokes Alg.~\ref{alg:da} at the line 2, and explores the Boolean structures of $\varphi_1$ and $\varphi_2$ at most two times,
ensuring the efficiency of the first stage.

The divergence analysis and isomorphism analysis over and under-approximate the equivalence, respectively.
Although the analyses do not always determine the equivalence,
they can handle a large proportion of data constraints in practice, evidenced by our evaluation.
Formally, we state two theorems to formulate the theoretical guarantee,
of which the proofs are provided in the appendices.

\begin{theorem}
	\label{thm:PSolvable}
	The steps in Alg.~\ref{alg:eqr} before line 6 run in polynomial time to $N$,
	where $N$ is the upper bound of the numbers of AST nodes for the two data constraints.
\end{theorem}

\begin{theorem}
	\label{thm:soundcomplete}
	For the syntax in Fig.~\ref{lang-syntax},
	the data constraints are semantically equivalent if Alg.~\ref{alg:eqr} returns true
	and vice versa.
\end{theorem}


\section{Implementation}
\label{sec:implementation}
We have implemented \ToolName\ in Python and deployed it in the FinTech company A.
\ToolName\  first generates the AST of a data constraint and then translates it to the symbolic representation. 
We leverage the Z3 SMT solver~\cite{MouraB08} to support the SMT solving in the third stage.
Particularly, we utilize the bit-vector, floating-point arithmetic, and string theory 
to encode variables and literals in the finite-length integer, floating point, and string types, respectively.

Based on \ToolName, 
we have further implemented two bots,
which are shown in Fig.~\ref{fig:app},
to conduct the equivalence clustering and searching, respectively.
In the equivalence clustering, we verify the equivalence of data constraints
by invoking \ToolName\ pairwise.
Particularly, we cache the symbolic representation of each data constraint to avoid redundant construction in different invocations.
Similarly, we examine the equivalence of a new data constraint and each existing one sequentially in the equivalence searching,
and also generate the symbolic representation for a data constraint only once.

\section{Evaluation}
\label{sec:eval}
To quantify the effectiveness and efficiency, 
we evaluate \ToolName\ upon the data constraints in a FinTech system by investigating the following research questions:
\begin{itemize}[leftmargin=*]
	\item \textbf{(RQ1)} How many equivalent data constraints are identified?
	\item \textbf{(RQ2)} How efficient is \ToolName\ in the equivalence clustering and searching?
	\item \textbf{(RQ3)} How important is each of the three stages?
\end{itemize}

\textbf{Subjects.}
We collect \TotalRuleNumber\ data constraints from a FinTech system in Company A,
which are in the syntax shown in Fig.~\ref{lang-syntax}.
Averagely, a data constraint contains 9.4 data constraints and 17.6 lines of code.
Despite the moderate average size,
we still need to handle the large set of data constraints efficiently,
which is non-trivial yet crucial in industrial scenarios.
Lastly, there are 1,497 data constraints not obeying our syntax, which are not selected as the subjects.
They mainly contain advanced string operations, e.g., \textit{substring} and \textit{replaceAll}, 
and system calls, e.g., \textit{getTimeZone}.
In this work, \ToolName\ focuses on the data constraints in our given syntax,
covering most of the data constraints (95.4\%) in the FinTech system of Company A.

\smallskip
\textbf{Environment.}
We conduct all the experiments on a 64-bit machine with 40 Intel(R) Xeon(R) CPU E5-2698 v4 @ 2.20 GHz and 512 GB of physical memory.
We invoke the Z3 SMT solver with its default options.
We run the experiments with a time limit of 12 hours and a memory limit of 16 GB.

\smallskip
\textbf{Availability.}
We release the code and sample constraints in GitHub repository~\cite{EqDACTool}.
The whole set of data constraints cannot be shared because of confidentiality agreements.

\subsection{Equivalent Data Constraint Identification}
\label{subsec:rq1}
To answer the first question,
we evaluate \ToolName\ upon \TotalRuleNumber\ data constraints by verifying the data constraint equivalence pairwise.
Specifically, each pair of data constraints is fed to \ToolName\ to determine whether they are equivalent.

\begin{figure}[t]
	\centering
	\includegraphics[width=\linewidth]{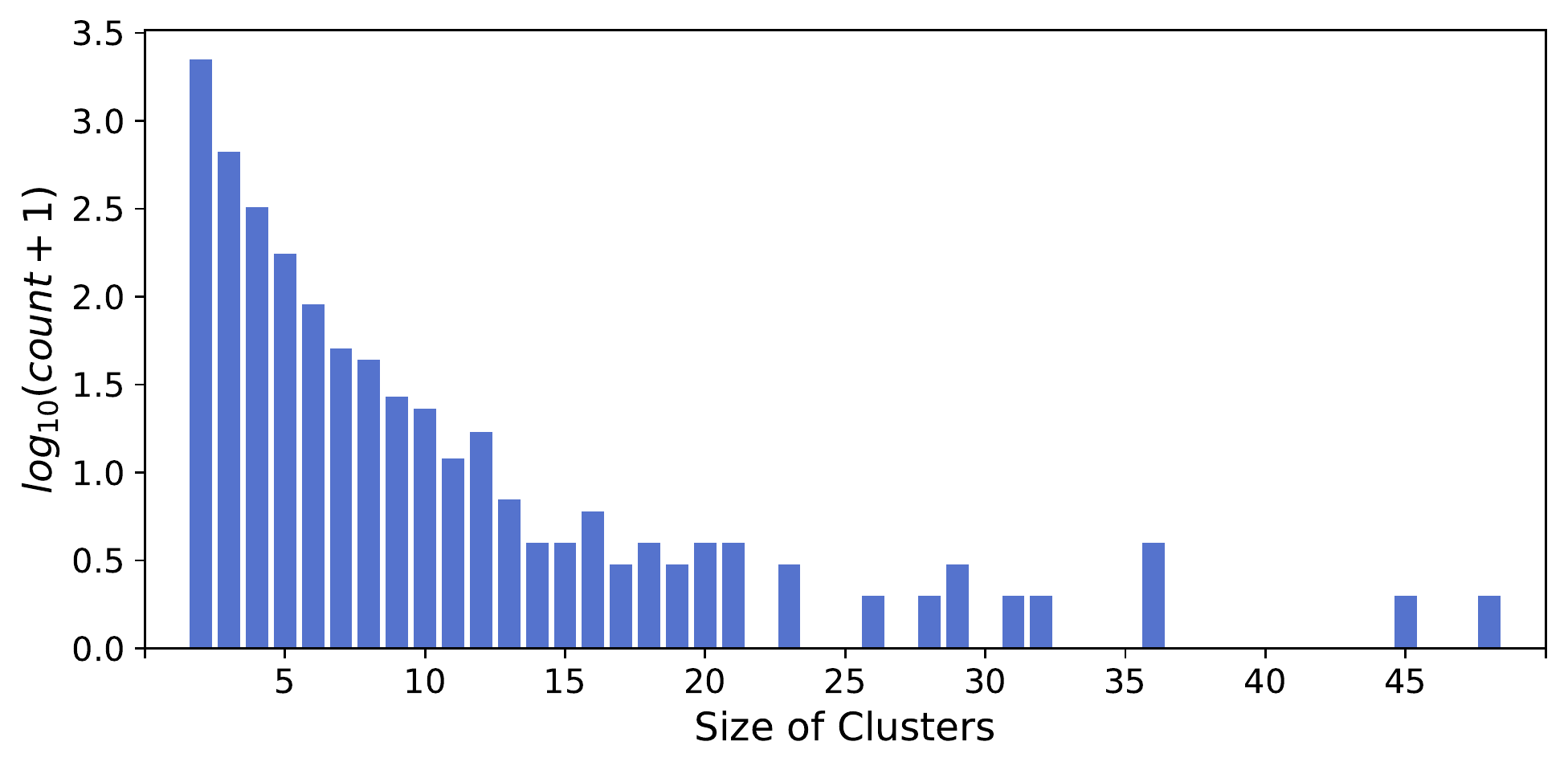}
	\caption{The counts and sizes of clusters}
	\label{eval:aggregation}
\end{figure}
\begin{figure}[t]
	\centering
	\includegraphics[width=\linewidth]{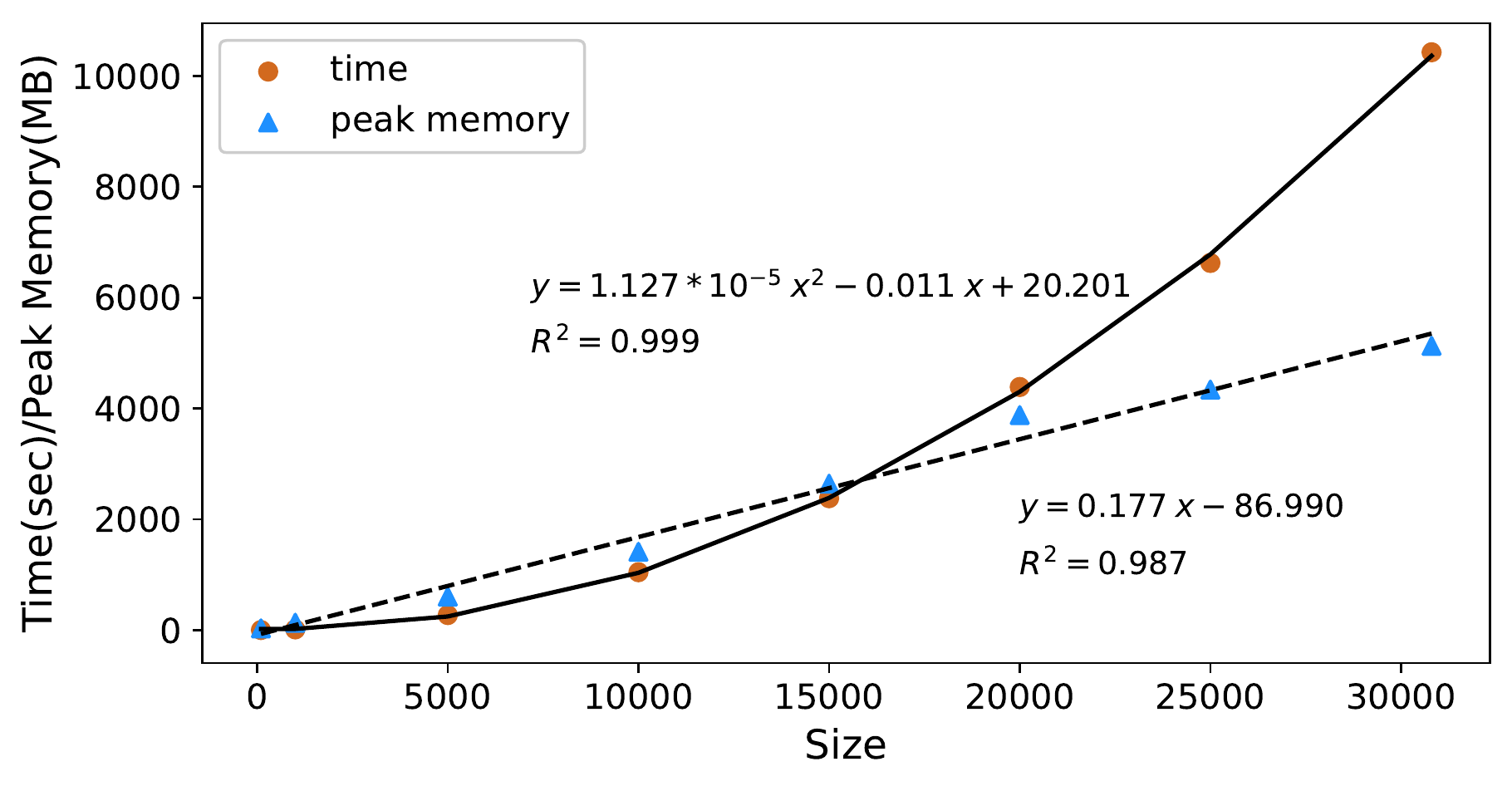}
	\caption{Time and memory cost of equivalence clustering}
	\label{eval:overhead1}
\end{figure}

\smallskip
\textbf{Result.}
We find that \TotalEqRuleNumber\ data constraints (37.5\%) have one or more equivalent variants,
forming \TotalEquivalentPairNumber\ equivalent pairs
and 3,696 equivalence clusters.
Particularly, we can leave one data constraint in each equivalence cluster 
and eliminate 7,842 data constraints without compromising the validity of the data reconciliation.
Due to our limited permission, we sample a subset of the data constraints and measure the CPU time reduction 
when avoiding checking redundant ones.
The result shows that the CPU time reduction ratio reaches 15.48\%.
According to the feedback of the experts,
any reduction can bring a drastic benefit to the overall system in the long run,
as data constraints are frequently checked online during a long development cycle.


We also count the data constraints in each cluster, 
in which data constraints are equivalent to each other.
Fig.~\ref{eval:aggregation} shows that the size of a cluster ranges from 2 to 48.
Specifically, the number of clusters with a size of 2 is 2,233.
For the largest cluster with the size of 48, a violation of any data constraint will generate 48 alerts. 
Therefore, identifying equivalent data constraints can provide practical guidance in reusing the checking results and support the redundant alert elimination.

\begin{figure*}[t]
\begin{minipage}{.71\textwidth}
\small
\centering
\subcaptionbox{\label{1}}{\includegraphics[width = .325\linewidth]{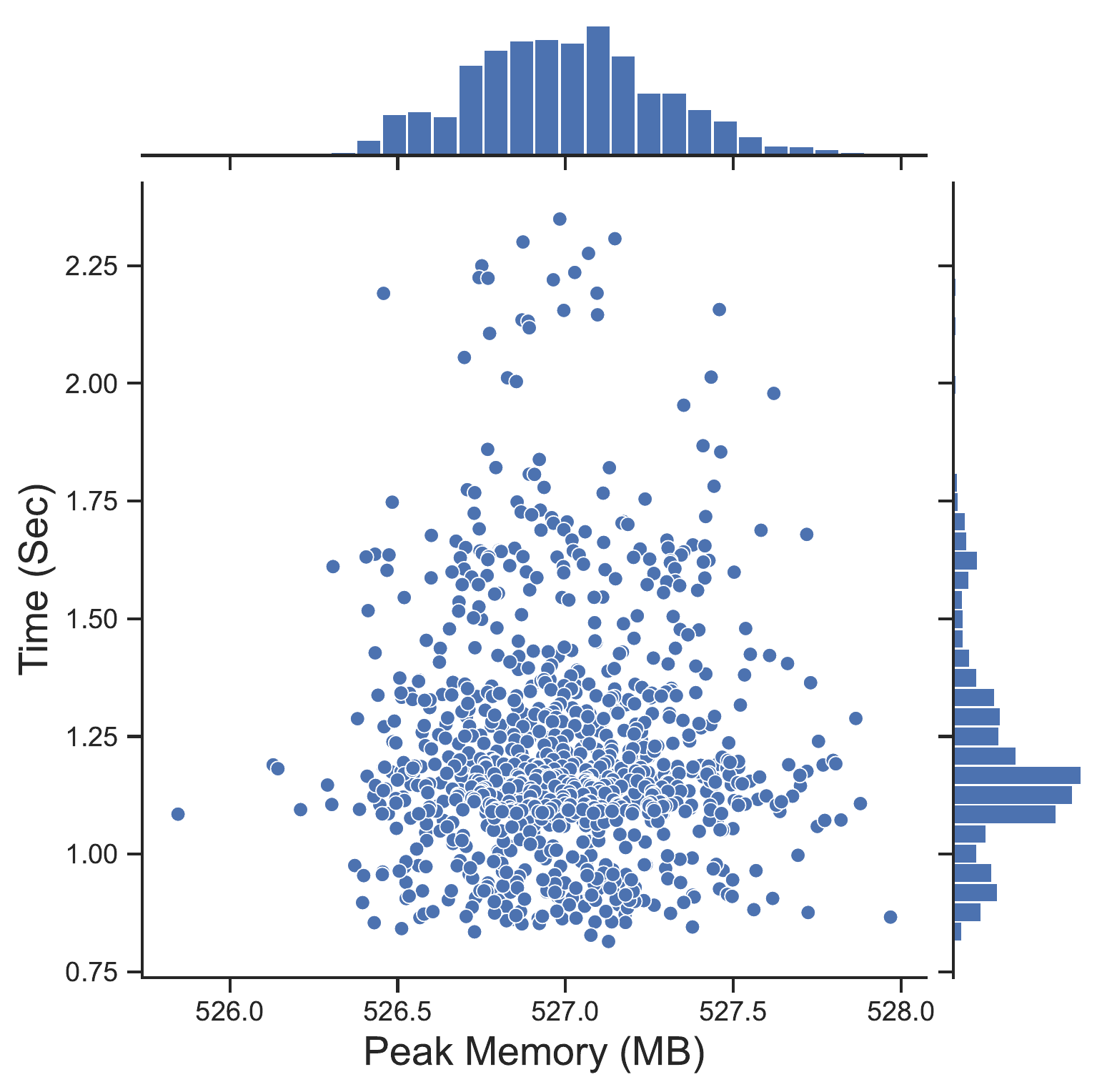}}
\subcaptionbox{\label{2}}{\includegraphics[width = .325\linewidth]{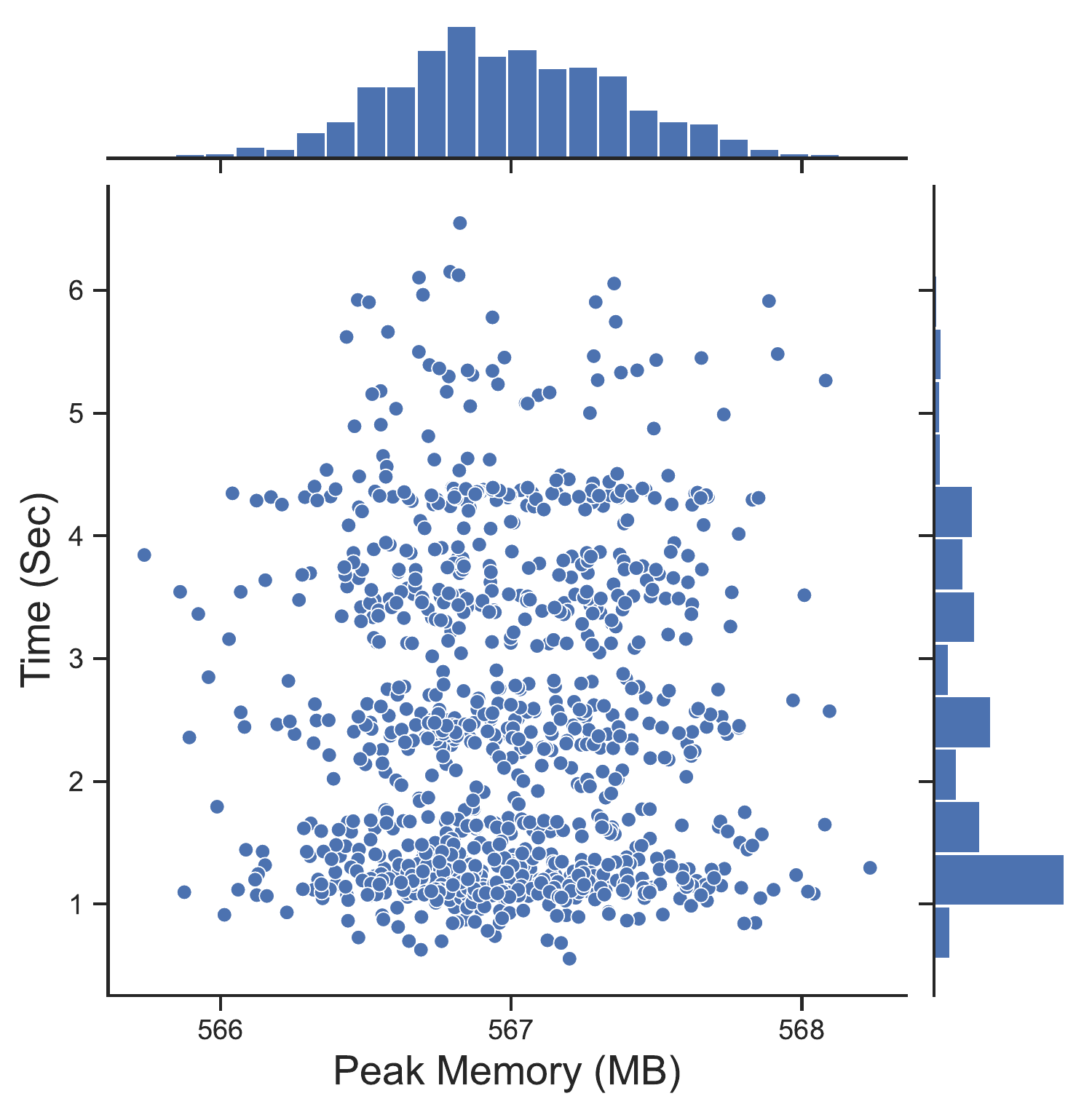}}
\subcaptionbox{\label{3}}{\includegraphics[width = .325\linewidth]{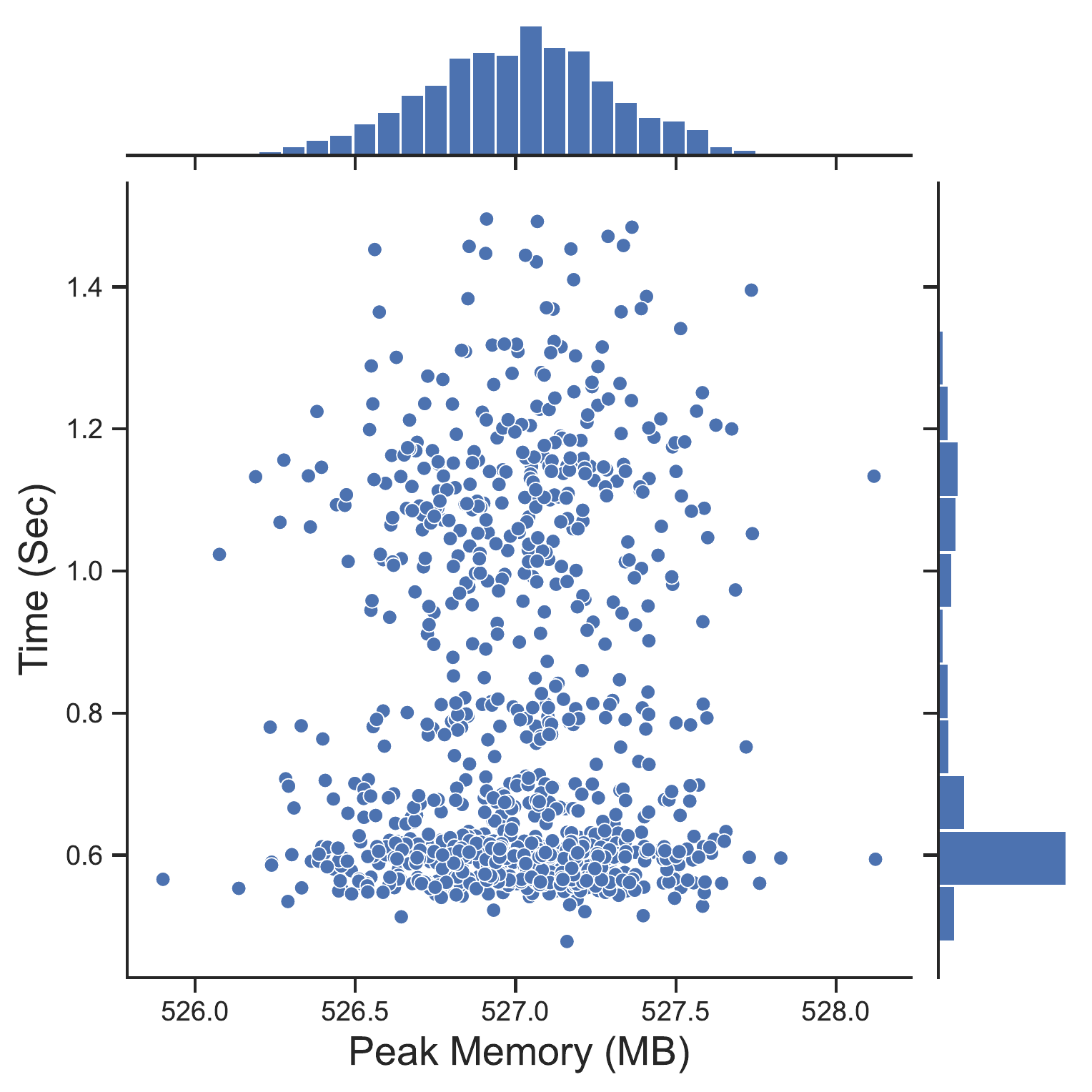}}
\caption{Time and memory cost of \ToolName, \ToolNameNoCertificate, and \ToolNameNoSemantics}
\label{fig:ablation_filtering}
\end{minipage}\hfill\vline\hfill%
\begin{minipage}{.27\textwidth}
\small
\centering
\begin{lstlisting}[style=mystyle]
  /* Data contraint 1 */
  assert(t.id != t.pid);
  assert(ut.oid != ut.iid);
  if(t.id == ut.oid){
  	assert(t.pid == ut.iid);
  } else {
    assert(t.id == ut.iid);
    assert(t.pid == ut.oid);}

  /* Data contraint 2 */
  if(t.id == ut.iid){ 
  	assert(ut.oid == t.pid);
  } else {
    assert(t.id == ut.oid);
    assert(t.pid == ut.iid);}
  assert(ut.iid != ut.oid);
  assert(t.pid != t.id);
\end{lstlisting}
\vspace{-2mm}
\caption{An example of case study}
\label{fig:example}
\end{minipage}
\end{figure*}

\subsection{Performance Evaluation}
\label{subsec:rq2}
We investigate the time consumption and memory usage of \ToolName\ in the equivalence clustering and searching.
The experimental configurations are set up as follows.
\begin{itemize}[leftmargin=*]
	\item \textit{Equivalence clustering}: 
	To quantify the cost of clustering different sizes of data constraint sets,
	we construct eight sets of the data constraints, of which the sizes range from 100 to \TotalRuleNumber,
	and measure the time and memory usage of the clustering. 
	All the data constraints are selected randomly.
	\item \textit{Equivalence searching}: 
	We select 1,000 data constraints from \TotalRuleNumber\ data constraints as the recently-submitted ones and regard the remaining as the existing ones.
	Specifically, half of the selected ones are equivalent to at least one data constraint in the remaining set to quantify the cost of the equivalence searching in the worst case.
\end{itemize}


\textbf{Result.}
As shown by Fig.~\ref{eval:overhead1},
\ToolName\ finishes analyzing \TotalRuleNumber\ data constraints in 2.89 hours within 5.01 GB of peak memory.
We perform the regression analysis to quantify the scalability,
choosing the quadratic and linear functions as the templates of the regression analyses
for the time and memory cost, respectively,
as we construct a symbolic representation for each data constraint only once and invoke the decision procedure in a pairwise manner.
The $R$-squared values for memory and time are 0.987 and 0.999, respectively.
Also, the coefficients in the quadratic and linear terms are quite small, indicating that the overhead increases gently.
In summary, \ToolName\ supports the scalable equivalence clustering.


Fig.~\ref{1} shows the cost of the equivalence searching.
All the analyses finish in 2.5 seconds within 528 MB of peak memory.
Specifically, there is little difference in memory cost,
ranging from 525.85 MB to 527.87 MB,
while the time cost has a relatively large variance.
The analyses of several data constraints demand SMT solving,
which introduces more time costs.
Typically, most of the cases can be analyzed in 1.5 seconds, 
and the average time cost is only 1.22 seconds.
Thus, \ToolName\ supports searching equivalent data constraints efficiently,
which is essential for maintenance.


\subsection{Ablation Study}
\label{subsec:rq3}
We set three ablations,
namely \ToolNameNoInvariant, \ToolNameNoCertificate, and \ToolNameNoSemantics,
which skip the divergence analysis, the isomorphism analysis, and SMT solving, respectively.
The first two ablations are sound and complete, 
while \ToolNameNoSemantics\ can return \emph{unknown} due to its incompleteness.

\begin{table}[t]
	\centering
	\caption{The statistics of the equivalence clustering}
	\resizebox{\linewidth}{!} 
	{
		\begin{tabular}{llccc}
			\toprule
			\textbf{Variant}  & \textbf{Time(h)} & \textbf{Mem(GB)} & \textbf{\#Eq Pair} & \textbf{\#Redundant}          \\ \midrule
			\ToolNameNoInvariant & OOT & 7.27 & 141 & 53\\
			\ToolNameNoCertificate & \NoCertificateTime & 6.80  &\TotalEquivalentPairNumber & \TotalDeletedRuleNumber\\
			\ToolNameNoSemantics & \NoSemanticsTime & 3.94 & \TotalMustEquivalentPairNumber & \TotalMustDeletedRuleNumber\\
			\ToolName &\TotalTime  & 5.01 & \TotalEquivalentPairNumber & \TotalDeletedRuleNumber\\
			\bottomrule
		\end{tabular}
	}
	\label{table:ablation_consolidation}
\end{table}

\smallskip
\textbf{Result.}
Table~\ref{table:ablation_consolidation} shows the results of the ablation study in the equivalence clustering.
As we can see, \ToolNameNoInvariant\ does not finish analyzing \TotalRuleNumber\ data constraints in 12 hours, 
and its peak memory reaches nearly 7.27 GB.
Specifically, \ToolNameNoInvariant\ only finishes comparing seven data constraints with the remaining data constraints pairwise,
discovering 141 equivalent pairs and 53 redundant data constraints.
\ToolNameNoCertificate\ discovers the same equivalent pairs as \ToolName.
However, it has to perform the SMT solving for all the data constraint pairs of which the equivalence is not refuted by the divergence analysis,
increasing the time cost to 4.48 hours.
\ToolNameNoSemantics\ skips the SMT solving and consumes less time and memory than \ToolName,
spending 1.78 and 0.35 hours on the divergence analysis and the isomorphism analysis, respectively.
It does not discover \TotalLastStageEquivalentPairNumber\ equivalent pairs,
and thus, misses \TotalLastStageEquivalentNumber\ redundant constraints.
Particularly, our divergence analysis identifies 38,964 non-equivalent pairs
even if their symbolic representations have the same sets of data variables, literals, and operators.
Thus, the divergence analysis not only refutes the equivalence soundly 
but also provides the possibility of refuting more non-equivalent pairs.

Fig.~\ref{fig:ablation_filtering} shows the cost of the equivalence searching.
\ToolNameNoCertificate\ 
costs more in each equivalence searching task,
as the SMT solver 
consumes more resources to prove the equivalence.
Specifically, its average time cost is 2.53 seconds, and its peak memory reaches 566.98 MB.
In the worst case,
it takes 6.56 seconds to finish the equivalence searching of a data constraint,
degrading its usability in real-world production.
\ToolNameNoSemantics\ consumes less time because it does not invoke SMT solvers in all the cases.
However, it can not identify the equivalent variants for~\FilteringDiffNumber\ data constraints due to incompleteness.
\ToolNameNoInvariant\ does not finish the equivalence searching of 1,000 data constraints in the given time budget.
It has to invoke the SMT solver to prove the non-equivalence, making the overall time cost unacceptable.

\smallskip
\textbf{Case Study.}
Fig.~\ref{fig:example} shows an equivalent pair discovered via the SMT solving.
The data constraints both examine whether the IDs of the income and expense accounts match with the ones in the transaction.
Unfortunately, we can not deduce the equivalence from the parse trees of their symbolic representations.
Instead, we have to reason multiple assertions in a relational manner.
The two assertions in the sequencing are the premise of the equivalence of two $\textbf{ite}_s$ statements,
while it is beyond the ability of the isomorphism analysis.


\subsection{Discussion}
In what follows, we demonstrate the discussions on the feedback from the users, threats to validity, limitations, and future work.

\smallskip
\textbf{Feedback from the Users}.
\ToolName\ has been integrated into the production line of Company A, 
serving as the core building block of two bots in the CI/CD workflow.
To obtain the feedback of users,
we assigned the questionnaires to the developers and the quality assurance managers in the forum of the company,
which received rave reviews from users.
For example, a developer comments the search bot in the forum as follows, showing his appreciation for the instant response and useful results.

\smallskip
\mybox{\emph{``The search is so smooth! I had been expecting such an assistant for data constraint maintenance. The results are mostly fetched in just one or two seconds, assisting in merging data constraints.''}}

\smallskip
\textbf{Threats to Validity}.
A threat to validity is whether the way of producing data constraints affects our results.
Ineffective communication between developers could increase the number of equivalent data constraints,
as they are unaware of the data constraints submitted by others.
For a small FinTech system with only a few data constraints, 
the benefit of resolving redundancy could be less significant. 
\ToolName\ mainly targets systems with thousands of data constraints and shows excellent potential to improve their maintainability.

\smallskip
\textbf{Limitations and Future Work}.
First, our syntax excludes several string operations.
Theoretically, solving general string constraints is undecidable~\cite{chen2017decidable,KiezunGAGHE12,chen2019decision}, while the first two stages of \ToolName\ can still work in the presence of advanced string operations.
It would be interesting to reason more string operations
even though, according to our experience, they do not widely exist.
Second, \ToolName\ focuses on equivalence relations in this work.
It is meaningful to examine whether a data constraint subsumes others for consolidation~\cite{SousaDVDG14}.
Third, \ToolName\ currently analyzes data constraints in FinTech systems.
It would be promising to extend \ToolName\ to support light-weighted equivalence checking in other domains,
such as SQL queries~\cite{Zhou19VLDB} and database-backed programs~\cite{EmaniRBS16}.

\section{Related Work}

\subsection{Program Equivalence Checking}
There is an extensive body of research on program equivalence checking, 
which is a crucial building block in many clients, such as translation validation~\cite{Necula00, SewellMK13} and program synthesis~\cite{Schkufza0A13,SamakKR20,NandiWAWDGT20}.
One line of studies reduces equivalence checking to proving specific verification conditions,
such as 
\textit{relational verification}~\cite{Mora0RC18,trostanetski2017modular,unno2021constraint,SousaD16, Churchill19PLDI,10.1145/3062341.3062378}.
Similar approaches include using symbolic execution for loop-free programs~\cite{PersonDEP08, lahiri2012symdiff, SamakKR20,badihi2020ardiff,chen2019relational}.
Different from the existing efforts  that target sophisticated program constructs,
\ToolName\ focuses on a domain-specific language in FinTech systems,
pursuing efficiency over the capability of handling a flexible program syntax.

Another line of studies proves program equivalence via  \textit{term rewriting}~\cite{Necula00, GoldbergZB05, Chu17HoTTSQL, Chu18Eq}.
The effectiveness relies heavily on the quality of rewrite rules. 
First, they may sacrifice soundness or completeness if the rule set contains an incorrect rule or misses a right one~\cite{Nandi21RInf}.
Second, they may suffer from the phase ordering problem~\cite{Max21Egg} in the presence of a large number of rewrite rules.
To obtain better complexity,~\cite{GuilloudK22} restricts the form of rewrite rules,
and adopts tree isomorphism algorithms to check syntactic isomorphism.
\ToolName\ bears similarities to~\cite{GuilloudK22} in terms of proving the equivalence,
while we consider more program constructs in the isomorphism analysis, such as arithmetic operators and string predicates,
which promotes its capability in practice.

\subsection{SQL Query Equivalence Checking}
Verifying SQL query equivalence is an essential topic in both academia and industrial communities.
The state-of-the-art approaches focus on specific forms of SQL queries~\cite{KolaitisV98}
and apply either algebraic reasoning techniques~\cite{Chu17HoTTSQL, Chu17SQL, Chu18Eq} or symbolic reasoning~\cite{WangCB18, Zhou19VLDB, zhou2020symbolic} for equivalence verification.
Typically, \UDP~\cite{Chu18Eq} 
utilizes U-semiring to encode the bag semantics of SQL effectively
and checks the isomorphism between two algebraic structures.
However, it fails to handle advanced features, e.g., three-valued logic,
and suffers from the inefficient chase procedure in the isomorphism checking~\cite{FaginKMP03}.
In contrast, \EQUITAS~\cite{Zhou19VLDB} encodes the semantics with a FOL formula
and leverages a solver to determine the equivalence, 
handling more SQL features~\cite{Zhou21sigmod}
than \UDP.

In our work, \ToolName\ targets the equivalence of data constraints rather than SQL queries,
while it bears similarities with existing efforts in terms of technical designs.
Specifically, it abstracts away the orders of commutative constructs,
and leverages a solver to determine the equivalence of two FOL formulas.
Thus, it avoids the inefficient chase procedure and unleashes the power of SMT solving.

\subsection{SMT Solving Optimization}
There is a vast amount of literature on guiding SMT solving with program features.
One typical line of studies performs semantically equivalent transformations to reduce the overhead of SMT solving~\cite{XuLFMZLY21,Cadar15,PerryMZC17,BouchetCCDGHJMP20,ChenH21},
For example, \cite{PerryMZC17} uses contextual information to simplify array constraints, 
which transforms array operations with symbolic indexes to the ones only involving constant indexes.
Another line of the literature utilizes semantic information to optimize SMT solving algorithms~\cite{ChenH18,FanL022,shi2021path,shuai2021type}.
For instance, \cite{ChenH18, ChenH21} leverage control-flow information to guide the branching strategy in CDCL($T$)-style SMT solving. 
Similarly, our symbolic representations preserve program syntactic features, which supports proving the equivalence efficiently by the isomorphism analysis.

Apart from accelerating the solving process, 
caching the intermediate results, e.g., unsatisfiable core~\cite{MrazekJSLB17}
and syntactic features~\cite{VisserGD12},
can also avoid calling the solver.
\Green\ examines the syntactic equivalence of constraints,
and reuses the previous solving result of the equivalent ones~\cite{VisserGD12}.
It shares a similar idea with the isomorphism analysis of \ToolName,
while \ToolName\ also conducts the divergence analysis to discover non-equivalent pairs efficiently,
promoting the efficiency of data constraint maintenance applications in the real world.

\section{Conclusion}
We have presented \ToolName, an efficient, sound, and complete decision procedure for verifying the data constraint equivalence in FinTech systems.
It supports two typical clients,
namely equivalence clustering and searching,
in the production line of a global FinTech company.
\ToolName\ scales to a large number of data constraints with high efficiency,
liberating productivity in the data constraint maintenance.
We believe that the insight behind \ToolName\ can further promote equivalence checking in other domains.

\section*{Acknowledgment}
We thank the anonymous reviewers for valuable feedback on earlier drafts of this paper, 
which helped improve its presentation. 
We also appreciate Dr. Xiao Xiao for insightful discussions.
The authors are supported by the RGC16206517, ITS/440/18FP and PRP/004/21FX grants from the Hong Kong Research Grant Council and the Innovation and Technology Commission, Ant Group, and the donations from Microsoft and Huawei.
Peisen Yao is the corresponding author.

\bibliographystyle{unsrt}
\bibliography{sigproc}

\section*{Appendices}

In this supplementary material, we provide the proofs of the theorems in our paper.

\subsection{Complexity Barrier}
In what follows, we present the proof of Theorem~\ref{thm:decomp}.

\begin{proof}
	We only need to prove that we can reduce any instance of the SAT problem to an instance of the data constraint equivalence problem in polynomial time.
	
	Consider an arbitrary propositional logic formula $\psi$, which contains $n$ variables denoted by $a_i$
	$(1 \leq i \leq n)$.
	We first construct a database table with $n$ attributes, namely $v_i$, where $1 \leq i \leq n$.
	Meanwhile, we introduce a constant set containing $n$ unique constants, denoted by 
	$L=\{\ell_1, \ell_2, \cdots, \ell_n\}$.
	We then construct the assertion statement $\textbf{assert}(f(v_1, v_2, \cdots, v_n))$.
	Here the boolean expression $f(v_1, v_2, \cdots, v_n)$ is constructed by replacing $a_i$ with the equality constraint $v_i == \ell_i$ in $\psi$.
	Such the assertion statement is exactly the data constraint $r$ we want.
	Obviously, $\psi$ is unsatisfiable if and only if $r$ is semantically equivalent to 
	$$\textbf{assert}(f(v_1, v_2, \cdots, v_n) \ \textbf{and} \ \lnot f(v_1, v_2, \cdots, v_n))$$
	The reduction can be achieved in linear time to the size of the formula $\psi$.
	Therefore, the data constraint equivalence problem is NP-hard.
\end{proof}

\subsection{Complexity}
Before proving Theorem~\ref{thm:PSolvable}, we first propose and prove two lemmas as follows.

\smallskip
\begin{lemma}
	\label{lemma1}
	We define the size of a symbolic condition $\phi$ as follows:
	\begin{equation*}
		\delta(\phi)=\left\{
		\begin{array}{cl}
			1 & \phi \ \text{is atomic}\\
			1 + \delta(\phi_0) & \phi = \lnot \phi_0\\
			\delta(\phi_1) + \delta(\phi_2) & \phi = \phi_1 \lor \phi_2 \text{ or } \phi = \phi_1 \land \phi_2
		\end{array} \right.
	\end{equation*}
	Given any data constraint $r$, denote the node number of its abstract syntax tree by $N$.
	The size of its symbolic representation $\delta(\varphi)$ is polynomial to $N$.
\end{lemma}

\smallskip
\begin{proof}
	For clarity, we introduce two functions $\alpha$ and $\beta$:
	\begin{itemize}[leftmargin=*]
		\item $\alpha(\mathbf{E}, e)$ is the number of terms that $e$ may be equal to.
		\item $\beta(\mathbf{E}, e)$ is the maximal size of the symbolic condition under which $e$ is equal to a specific term.
		i.e., $\beta(\mathbf{E}, e) = \max_{(\tau, \phi)\in \mathbf{E}(e)} \delta(\phi)$.
	\end{itemize}
	
	According to the rules in Fig.~\ref{rule:statement} and Fig.~\ref{rule:boolexpr},
	$\mathbf{E}$ is only updated by the rules \texttt{ASSIGN}, \texttt{SEQ}, and \texttt{ITE-S}.
	
	\begin{itemize}
		\item Let's consider $\alpha(\mathbf{E}', v)$ and $\beta(\mathbf{E}', v)$ after applying the rule \texttt{ASSIGN}.
		If the rule \texttt{ASSIGN} applies the rule \texttt{VAR}, we have
		$$\alpha(\mathbf{E}', v) = 1, \ \ \beta(\mathbf{E}', v) = 1$$
		If the rule \texttt{ASSIGN} applies the rule \texttt{AE}, we have
		$$\alpha(\mathbf{E}', v) = 1, \ \ \beta(\mathbf{E}', v) = \beta(\mathbf{E}, a_1) + \beta(\mathbf{E}, a_2)$$

		\item After applying the rule \texttt{ITE-S}, for any  $e \in dom(\mathbf{E}')$, we have
		$$\alpha(\mathbf{E}', e) = O(\alpha(\mathbf{E}_1, e) + \alpha(\mathbf{E}_2, e))$$
		$$\beta(\mathbf{E}', e) = \max_{i \in \{1, 2\}} \beta(\mathbf{E}_i, e) = O(\beta(\mathbf{E}_1, e) + \beta(\mathbf{E}_2, e))$$
		
		\item After applying the rule \texttt{SEQ}, the effects of the involved rules accumulate.
	\end{itemize}
	
	Based on the above equations, we can find that $\alpha(\mathbf{E}, e)$ and $\beta(\mathbf{E}, e)$
	are both linear to the times of applying the rules.
	Thus, for any program location, we have
	$$\max_{e \text{ in } r} \alpha(\mathbf{E}, e) = O(N), \ \ \max_{e \text{ in } r} \beta(\mathbf{E}, e) = O(N)$$
	
	Now, we can estimate the upper bound of $\delta(\varphi)$.
	According to the rules in Fig.~\ref{rule:statement} and Fig.~\ref{rule:boolexpr},
	$\mathbf{\Phi}$ is only updated by the rules \texttt{ASSERT}, \texttt{ITE-S}, and \texttt{SEQ}.
	\begin{itemize}
		\item Let's consider $\delta(\mathbf{\Phi}')$ after applying the rule \texttt{ASSERT}.
		If the rule \texttt{ASSERT} applies the rule \texttt{ACmp}, we have
		$$\delta(\mathbf{\Phi}')-\delta(\mathbf{\Phi}) = \sum_{(t_i, \phi_i) \in \mathbf{E}_i(u_i)} (1 + \delta(\phi_1) + \delta(\phi_2)) = O(N^2)$$
		If the rule \texttt{ASSERT} applies the rule \texttt{ITE-E}, we have
		$$\delta(\mathbf{\Phi}')-\delta(\mathbf{\Phi}) = \delta(\gamma_1) + \delta(\gamma_2) + 3$$
		Observe that $\gamma_1$ and $\gamma_2$ are obtained by applying the rules \texttt{ACmp} and \texttt{ITE-E}.
		We can sum up the above two equations and obtain that
		$$\delta(\mathbf{\Phi}')-\delta(\mathbf{\Phi}) = O(N^3)$$
		
		\item After applying the rule \texttt{ITE-S}, we have the following relation:
		$$\delta(\mathbf{\Phi}') = 3 + \delta(\mathbf{\Phi}_1) + \delta(\mathbf{\Phi}_2)$$
		
		\item The rule \texttt{SEQ} accumulates the effects upon $\delta(\mathbf{\Phi})$.
	\end{itemize}
	
	By summating all the above equations, we have
	$$\delta(\mathbf{\Phi}_e) = O(N^4)$$
	Notice that the negation elimination can reduce the size of the FOL formula.
	Therefore, we have
	$$\delta(\varphi) \leq \delta(\mathbf{\Phi}_e) = O(N^4)$$
	
	Notably, the estimated upper bounds of $\max_{e \text{ in } r} \alpha(\mathbf{E}, e)$, $\max_{e \text{ in } r} \beta(\mathbf{E}, e)$, and $\delta(\mathbf{\Phi}_e)$ are not tight.
	For clarify, we only attempt to bound them with the polynomial function of $N$.
	The upper bounds can be further strengthened by the polynomial functions of the numbers of specific AST nodes.
\end{proof}

\smallskip
\begin{lemma}
	\label{lemma2}
	Given a data constraint $r$, its symbolic representation $\varphi$ can be constructed in polynomial time to $N$,
	where $N$ is the node number of the abstract syntax tree of $r$.
\end{lemma}

\smallskip
\begin{proof}
	We examine the time complexity of the rules in Fig.~\ref{rule:statement} and Fig.~\ref{rule:boolexpr}.
	The rules \texttt{ACmp} and \texttt{AE} can be applied in $O(N^2)$ time,
	as they have to iterate two sets pairwise.
	The other rules are all applied in $O(1)$ time, 
	as they only need to construct a constant number of FOL formulas.
	Each of the above rules is applied at most $O(N)$ times.
	Therefore, the symbolic representation can finally be constructed in $O(N^3)$.
\end{proof}

Now, we present the proof of Theorem~\ref{thm:PSolvable} as follows.
\begin{proof}
	We prove the theorem by proving three parts.
	\begin{itemize}[leftmargin=*]
		\item First, we can obtain that $\varphi_1$ and $\varphi_2$ can be constructed in polynomial time to $N$ based on Lemma~\ref{lemma2}.
		\item Second, the divergence analysis actually traverses the parse trees of $\varphi_1$ and $\varphi_2$,
		of which the sizes are both polynomial to $N$,
		as Lemma~\ref{lemma1} indicates that $\delta(\varphi_1)$ and $\delta(\varphi_2)$ are polynomial to $N$.
		Thus, the divergence analysis also works in polynomial time.
		\item Third, the function \textit{SCTree} constructs the parse trees in $O(\delta(\varphi_1) + \delta(\varphi_2))$ time, 
		which is polynomial to $N$.
		The AHU algorithm also works in $O(M)$ time, where $M$ is the node number of the tree.
		According to Lemma~\ref{lemma1}, $M$ is polynomial to $N$.
		Thus, the isomorphism analysis works in polynomial time.
	\end{itemize}
	Therefore, the steps in Alg.~\ref{alg:eqr} before line 6 run in polynomial time to $N$.
\end{proof}

\smallskip
At the end of the section, we want to emphasize the following two points.
First, we omit the discussion of several rules that are not shown in Fig.~\ref{rule:statement} and Fig.~\ref{rule:boolexpr} when proving the two lemmas,
e.g., the rules of evaluating the boolean expressions with logical connectives.
However, the arguments are similar to the rules that are discussed in the proofs.
Actually, the two lemmas hold for any data constraints in the syntax shown in Fig.~\ref{lang-syntax}.
Second, we do not provide the tight estimation of the complexity.
As shown in the proof of Lemma~\ref{lemma1},
the upper bound of $\delta(\mathbf{\Phi}_e)$ would be quite sophisticated,
involving with the numbers of program constructs in different kinds,
if we want to give a tight bound.
In this work, we only tend to show that the steps before the SMT solving can be achieved in polynomial time,
while the SMT solving may consume exponential time cost theoretically.

\subsection{Soundness and Completeness}
In this section, we prove the soundness and completeness.

\smallskip
\begin{proof}
	If two data constraints $r_1$ and $r_2$ are semantically equivalent,
	the divergence analysis does not return \emph{true},
	as it can not find an interpretation making their symbolic representations $\varphi_1$ and $\varphi_2$ evaluate to different truth values.
	Meanwhile, $\varphi_1$ and $\varphi_2$ are essentially the FOL formulas in the fragment
	of bit-vector theory, floating-point arithmetic theory, and word equations,
	which is theoretically decidable.
	Therefore, the SMT solving must terminate and return \emph{UNSAT}, making Alg.~\ref{alg:eqr} returns \emph{true}.
	
	\smallskip
	If Alg.~\ref{alg:eqr} returns \emph{true},
	the isomorphism analysis returns \emph{true} or the SMT solving returns \emph{UNSAT}.
	In the first case, the parse trees of two symbolic representations $\varphi_1$ and $\varphi_2$ are isomorphic,
	indicating that they must evaluate to the same truth values under a given interpretation,
	so the data constraints are semantically equivalent.
	In the second case, we have $\varphi_1$ and $\varphi_2$ are logically equivalent,
	implying the data constraint equivalence.
	Therefore, \ToolName\ is sound and complete.
\end{proof}

\end{document}